\documentclass[a4paper,12pt]{article}
\usepackage{amsthm,amsmath,amssymb,amsfonts, graphicx}

\def \AU {\, \mbox{Aut}\,}
\def \AU0 {\, \mbox{Aut}_0\,}
\def \id {\, \mbox{id}}
\def \ba {{\mathbf{a}}}
\def\bbbn{{\mathbb N}}
\def\bbbc{{\mathbb C}}
\def\bbbz{{\mathbb Z}}

\def\bbbd{{\mathbb D}}
\def\bbbt{{\mathbb T}}
\def\bbbo{{\mathbb O}}
\def\bbbi{{\mathbb I}}

\def\cb{{\bar{\mathbb C}}}

\def\cA{{\cal A}}
\def\cB{{\cal B}}

\def\cR{{\cal R}}
\def\cG{{\cal G}}

\def\cJ{{\cal J}}

\def\cP{{\cal P}}
\def\cD{{\cal D}}
\def\cV{{\cal V}}
\def\cW{{\cal W}}

\def\gA{{\mathfrak A}}

\def\bbe{{\bf e}}
\def\bbf{{\bf f}}
\def\bbh{{\bf h}}
\def\bbs{{\bf s}}
\def\bba{{\bf a}}
\def\bbb{{\bf b}}

\newtheorem{Def}{Definition}
\newtheorem{The}{Theorem}
\newtheorem{Pro}{Proposition}
\newtheorem{Lem}{Lemma}

\begin{document}
\bibliographystyle{unsrt}
\title{Automorphic Lie algebras and corresponding integrable systems. }

\author{Rhys T. Bury  and  Alexander V. Mikhailov 
\\
Applied Mathematics Department, University of Leeds, UK 
}

\maketitle

\begin{abstract} 
We  study automorphic Lie algebras and their applications to integrable systems. Automorphic Lie algebras are a natural generalisation of celebrated Kac-Moody algebras to the case when the group of automorphisms is not cyclic. They are infinite dimensional and almost graded. We formulate the concept of a graded isomorphism and classify $sl(2,\bbbc)$ based automorphic Lie algebras corresponding to all finite reduction groups. We show that hierarchies of integrable systems, their Lax representations and master symmetries can be naturally formulated in terms of automorphic Lie algebras.   

\end{abstract}

\section{Introduction}

The integrability of a nonlinear partial differential or a differential
difference
equation can often be related to the existence of a corresponding Lax
representation.
Having a Lax operator we can construct an infinite hierarchy of commuting
symmetries, local conservation laws and find exact multi-soliton solutions. It
enables us to find a recursion operator and a multi-Hamiltonian structure for
the corresponding 
equation. Symmetries, local conservation laws, recursion operators and
multi-Hamiltonian structures are fundamental properties of integrable equations
\cite{mr91k:58005}, \cite{integrability}. Integration of such equations can be reduced to
a
direct and inverse spectral transform associated with the Lax operator. 

Symmetries of the Lax operator play a key role in the spectral transform and
are reflected in all structures associated with the corresponding integrable
equation. Discrete groups of automorphisms of Lax operators, the {\sl reduction
groups}, were introduced in  \cite{mik79}, \cite{mik80}, \cite{mik81}. Reduction groups
have been extensively applied for the construction of new integrable systems,
recursion operators, $R$ matrices,  for the classification of soliton
solutions, and the spectral theory of Lax operators (see for example 
\cite{mik80}, \cite{mik81}, \cite{Mikhailov198251}, 
\cite{mik_dis}, \cite{LM04}, \cite{ wang09}, \cite{bmw2017}). 

Often the structure of Lax operators have a natural Lie algebraic interpretation in
terms of Kac-Moody algebras \cite{mr86h:58071}. A new class of Lie algebras over rings of automorphic functions, 
which can be also regarded as infinite dimensional Lie algebras over $\bbbc$, 
was proposed in \cite{mik_dis}. 
These  algebras  have been further studied in \cite{LM05} where they  
acquired the name {\sl automorphic Lie algebras} (see also \cite{L-thesis04}). Automorphic Lie algebras are  a natural generalisation
of Kac-Moody algebras. While a Kac-Moody algebra can be seen as a subalgebra
of a loop algebra, which is invariant with respect to a cyclic group of a finite order automorphism (the Coxeter automorphism \cite{mr86h:58071}), an automorphic Lie algebra is a subalgebra  of a generalised 
loop algebra which is
invariant with respect to a reduction group (the reduction group can be 
non-cyclic, 
noncommutative, and it can be infinite). 

Automorphic Lie 
algebras are infinite dimensional (over the field $\bbbc$), they are  almost graded
and can be characterised by a finite set of structure constants. They have a structure of a finitely generated $\bbbc[J]$--Lie module, where $J$ is a primitive automorphic function. The classification of automorphic Lie algebras is part of the
programme of classification of Lax operators and hence of integrable systems. The problem of classification of automorphic Lie algebras corresponding to finite reduction groups had been extensively studied in \cite{bury_thesis} and independently in \cite{ls10}. An alternative   approach  to automorphic Lie algebras and further development can be found in  \cite{knib}, \cite{klv}. Automorphic Lie algebras have found further applications to construction of differential-difference and partial-difference integrable systems and Yang-Baxter maps \cite{sotiris_mik2013}, \cite{bmx}.

In this paper we
define the concept of a graded isomorphism of almost graded algebras. It is stronger than isomorphism and can be effectively verified. We also study automorphic
Lie algebras related to the simple Lie algebra $A_1$ and finite reduction groups. 
We show that there
are five types of non-isomorphic algebras which include the polynomial part of
the $A_1$ loop algebra, the polynomial part of the
Kac-Moody algebra $A_1^1$ and three others. Explicit realisation of these algebras in terms of finitely generated $\bbbc[J]$--Lie modules is presented in Section \ref{explicit}. We discuss the construction of  Lax operators, corresponding
integrable hierarchies and master symmetries in terms of automorphic Lie algebras and illustrate it with examples.  
  
\section{Kac-Moody and automorphic Lie algebras}

The construction of automorphic Lie algebras is similar to the construction used 
in the theory of Kac-Moody 
Lie algebras. While a Kac-Moody algebra can be realised as a subalgebra of a 
loop algebra, which is invariant 
with respect to a cyclic group generated by an automorphism of a finite order, 
an automorphic Lie algebra can 
be viewed as a subalgebra of a simple Lie algebra over the field 
$\bbbc(\lambda)$, which is  invariant with 
respect to a finite group of automorphisms. Automorphic Lie algebras can also 
be defined for infinite groups, but 
in this paper we focus on the case of finite groups.

Let $\Gamma=\{\mu_k\in\bbbc\}$ denote a finite set of points and 
$\cR_{\lambda}(\Gamma)$ denote a ring of 
rational functions of the variable $\lambda$ with poles at $\lambda=\mu_k,\ 
\mu_k\in\Gamma$ and regular elsewhere. In this notation the ring of 
polynomials $\bbbc[\lambda]=\cR_{\lambda}(\infty)$ and the 
ring of Laurent polynomials 
$\bbbc[\lambda^{-1},\lambda]=\cR_{\lambda}(0,\infty)$.

Let $\gA$ be a simple Lie algebra over $\bbbc$ and
\begin{equation}\label{gAG}
\gA_{\lambda} (\Gamma)=\cR_{\lambda}(\Gamma)\otimes_\bbbc \gA\, . 
\end{equation}
Then $\gA_{\lambda} (\Gamma)$ may be made into a Lie algebra in a unique way 
satisfying 
\[ [p\otimes a,q\otimes b]=pq\otimes [a,b] \]
for $p,q\in \cR_{\lambda}(\Gamma),\ a,b\in \gA$. In particular, the algebra 
$\gA_{\lambda} (0,\infty)=
\cR_{\lambda}(0,\infty)\otimes_\bbbc \gA$ is called the {\sl loop algebra} 
\cite{carter}.
Elements $a(\lambda)\in\gA_{\lambda} (0,\infty)$ are  Laurent polynomials 
$\sum_{n\in\bbbz}\lambda^n a_n$
 where $a_n\in\gA$ with finitely many $a_n\ne 0$. We shall call the algebra 
$\gA_{\lambda} (\Gamma)$ a generalised loop algebra.

\subsection{Kac-Moody algebras}

Let $\phi_1 :\gA \rightarrow \gA$ be an automorphism of a finite order $n$, then 
$\Phi_1:\gA_{\lambda} (0,\infty)
\rightarrow \gA_{\lambda} (0,\infty)$, defined for any 
$a(\lambda)\in\gA_{\lambda} (0,\infty)$ as
\begin{equation}\label{autPhi}
 \Phi_1(a(\lambda))=\phi_1(a(\omega^{-1}\lambda))\, ,\qquad 
\omega=\exp(\frac{2\pi i}{n})
\end{equation}
is an automorphism of $\gA_{\lambda} (0,\infty)$. The automorphism $\Phi_1$ is 
of order $n$ and thus it generates a 
cyclic group of automorphisms  $\cG=\langle 
\Phi_1\,;\,\Phi_1^n=\id\,\rangle\simeq\bbbz/n\bbbz$. 

A Kac-Moody algebra $L(\gA,\phi_1)$ can be defined\footnote{There are many 
comprehensive monographs and textbooks 
presenting the theory of Kac-Moody algebras (see for example 
\cite{kac94}, \cite{carter}). We shall adopt definitions and 
some notations from \cite{mr86h:58071} (Section 5) where the most convenient 
and useful (for our purposes) exposition of 
Kac-Moody algebras is given. In \cite{mr86h:58071} and in this paper Kac-Moody 
algebras are assumed to be centreless, i.e. 
are quotients of the corresponding affine Lie algebras over their centres.} as 
a subalgebra of $\gA_{\lambda} (0,\infty)$ 
invariant with respect to the cyclic group of automorphisms $\cG$
\begin{equation}\label{KacMoody}
L(\gA,\phi_1)=\{a(\lambda )\in\gA_{\lambda} (0,\infty)\, |\, a( 
\lambda)=\phi_1(a(\omega^{-1}\lambda))\}\, \,\, .
\end{equation}
We have $L(\gA,\phi_1)=\sum_{k\in\bbbz}\lambda^k\gA_k$ where 
$\gA_k=\{a\in\gA\,|\, \phi_1(a)=\omega^k a\}$ and define  
$L^k(\gA,\phi_1)=\lambda^k\gA_k$. It is a graded 
Lie algebra
\[ L(\gA,\phi_1)=\bigoplus_{k\in\bbbz}L^k(\gA,\phi_1),\qquad 
[L^k(\gA,\phi_1),L^m(\gA,\phi_1)]\subset L^{k+m}(\gA,\phi_1)\, .\]

We can also consider two subalgebras $L_{\pm}(\gA,\phi_1)\subset L(\gA,\phi_1)$ 
of polynomials in $\lambda$ and $\lambda^{-1}$:
\[ L_+(\gA,\phi_1)=\{a\in\gA_{\lambda} (\infty)\, |\, a=\Phi_1(a)\},\quad 
L_-(\gA,\phi_1)=\{a\in\gA_{\lambda} (0)\, |\, a=\Phi_1(a)\}\, .
\]
The subalgebras $L_{\pm}(\gA,\phi_1)$ are isomorphic and they cover $L(\gA,\phi_1)$:
\begin{equation}
 L_-(\gA,\phi_1)\bigcup L_+(\gA,\phi_1)= L(\gA,\phi_1)\, , \quad  
L_-(\gA,\phi_1)\bigcap  L_+(\gA,\phi_1)=\gA_0\, .
\label{cover}
\end{equation}

{\bf Example:} 
In $\gA=sl(2,\bbbc)$ we take the standard (Cartan-Weyl) basis  ${\bf e,f,h}$ 
\[ {\bf e}= \left(\begin{array}{cc}
0&1\\0&0
\end{array}\right),\quad {\bf f}= \left(\begin{array}{cc}
0&0\\1&0
\end{array}\right),\quad {\bf h}= \left(\begin{array}{cc}
1&0\\0&-1
\end{array}\right) 
\]
with  commutation relations
\begin{equation}
 \label{ls2}
[\bbe,\bbf]=\bbh,\quad [\bbh,\bbe]=2\bbe,\quad [\bbh,\bbf]=-2\bbf\, .
\end{equation}
We define the automorphism $\Phi_1$ of order $2$ as
\begin{equation}\label{z2auto}
  \Phi_1(a(\lambda))=\left(\begin{array}{cc}
1&0\\0&-1
\end{array}\right)a(-\lambda)\left(\begin{array}{cc}
1&0\\0&-1
\end{array}\right)\, .
\end{equation}
Then $\gA_{2k-1}=\mbox{Span}_{\bbbc}\, (\bbe,\bbf), \ 
\gA_{2k}=\mbox{Span}_{\bbbc}(\bbh),\ k\in\bbbz$ and 
\begin{equation}\label{kacA1}
 L(\gA,\phi_1)=\bigoplus_{k\in\bbbz}\gA_k \lambda^k,\quad
L_{+}(\gA,\phi_1)=\bigoplus_{k\geqslant 0}\gA_k  \lambda^k,\quad
L_{-}(\gA,\phi_1)=\bigoplus_{k\leqslant 0}\gA_k  \lambda^k\, .
\end{equation}
The algebra $L(\gA,\phi_1)$ is isomorphic to the loop 
algebra $\gA_{\lambda}(0,\infty)$. Indeed, the set $\{e_k=\lambda^k\bbe,
f_k=\lambda^k\bbf,h_k=\lambda^k\bbh\}_{k\in\bbbz}$ is a basis in   
$\gA_{\lambda}(0,\infty)$ with non-vanishing commutation relations
\begin{equation}\label{loopcomm}
 [e_k,f_p]=h_{k+p},\quad [h_k,e_p]=2e_{k+p},\quad [h_k,f_p]=-2f_{k+p},\qquad 
k,p\in\bbbz. 
\end{equation}
In $L(\gA,\phi_1)$ one can take the basis $\{e^k=\lambda^{2k+1}\bbe,
f^k=\lambda^{2k-1}\bbf,h^k=\lambda^{2k}\bbh\}_{k\in\bbbz}$ and verify that the 
commutators of its elements are exactly the same as in (\ref{loopcomm}). 

This is an illustration of a general Theorem (V.Kac \cite{kac94}) that for
any simple Lie algebra $\gA$ and a finite order inner automorphism $\phi_1$ the
corresponding Kac-Moody algebra $L(\gA,\phi_1)$ is isomorphic to the loop
algebra $\gA_{\lambda}(0,\infty)$. Here we would like to stress the fact 
that 
the subalgebras $L_{\pm}(\gA,\phi_1)$ and $\gA_0$ in the coverage (\ref{cover}) depend on the choice of the automorphism and that is of importance to our
applications to integrable systems.

\subsection{Automorphic Lie algebras}

The map 
\begin{equation}
g_1:\cR_{\lambda}(0,\infty)\rightarrow\cR_{\lambda}(0,\infty),\quad 
g_1(\alpha(\lambda))=\alpha(\omega^{-1}\lambda), 
\quad \alpha(\lambda)\in\cR_{\lambda}(0,\infty),\quad \omega=\exp\left(\frac{2\pi i}{n}\right)
\end{equation}
 is an automorphism of order $n$ of the ring $\cR_{\lambda}(0,\infty) $.  The 
ring  $\cR_{\lambda}(0,\infty) $ has another 
automorphism $g_2$ of order $2$ 
\begin{equation}
g_2:\cR_{\lambda}(0,\infty)\rightarrow\cR_{\lambda}(0,\infty),\quad 
g_2(\alpha(\lambda))=\alpha(\lambda^{-1}), 
\quad \alpha(\lambda)\in\cR_{\lambda}(0,\infty)\, .
\end{equation}

The automorphisms $g_1$ and $g_2$ generate a subgroup $G\subset \mbox{Aut}\, 
\cR_{\lambda}(0,\infty)$ 
which is isomorphic to $\bbbd_n$ - the group of a dihedron with $n$ vertices. 
Indeed, we have $g_1^n=g_2^2=\id$ 
and it is easy to verify that $g_1g_2g_1g_2=\id$, thus
\begin{equation}
 \label{D2}
G=\langle g_1,g_2\, ;\, g_1^n= g_2^2=g_1g_2g_1g_2=\id\, \rangle\simeq \bbbd_n\, 
.
\end{equation}
The order $|\bbbd_n|=2n$. In the case $n=2$ the group is commutative and
$\bbbd_2 \simeq \bbbz/2\bbbz\times\bbbz/2\bbbz$ (the group of Klein).

The subring of all $G$--invariant (or automorphic) Laurent polynomials is given 
by
\[ 
  \cR^{G}_{\lambda}(0,\infty)=\{\alpha\in  \cR_{\lambda}(0,\infty)\,|\, 
g_1(\alpha)=g_2(\alpha)=\alpha\}.
\]
The ring $\cR^{G}_{\lambda}(0,\infty)=\bbbc[J]$, where 
$J=\frac{1}{2}(\lambda^n+\lambda^{-n})\in \cR_{\lambda}(0,\infty)$ 
is an automorphic Laurent polynomial. Moreover $J$ is a primitive automorphic 
function of the group $G$, in the sense that any 
automorphic rational function of $\lambda$ is a rational function of $J$ 
\cite{LM05}.

Let $\phi_1,\phi_2$ be two automorphisms of $\gA$ 
satisfying the conditions $\phi_1^n=\phi_2^2=\phi_1\phi_2\phi_1\phi_2=\id$. 
Then $\Phi_1$ (defined in (\ref{autPhi})) and $\Phi_2:\gA_{\lambda} 
(0,\infty)\rightarrow \gA_{\lambda} (0,\infty)$
\begin{equation}\label{autPhi2}
 \Phi_2(a(\lambda))=\phi_2(a(\lambda^{-1}))\, ,\qquad a(\lambda)\in\gA_{\lambda} 
(0,\infty)
\end{equation}
generate a subgroup $\cG=\langle 
\Phi_1,\Phi_2\,;\,\Phi_1^n=\Phi^2_2=\Phi_1\Phi_2\Phi_1\Phi_2=\id\,
\rangle\subset\mbox{Aut}\, 
\gA_{\lambda}(0,\infty)$ ({\sl a reduction group }
\cite{mik79}-\cite{mik_dis}), which is isomorphic to the dihedral group 
$\cG\simeq\bbbd_n$. The subalgebra of $ \gA_{\lambda}(0,\infty)$  
invariant with respect to the group of automorphisms $\cG$
\begin{equation}\label{sl2d2}
  \gA^{\cG}_{\lambda}(0,\infty)=\{a(\lambda )\in\gA_{\lambda} (0,\infty)\, |\, 
a=\Phi_1(a)=\Phi_2(a)\}\, \,
\end{equation}
is an example of an automorphic Lie algebra. In this example 
$\gA^{\cG}_{\lambda}(0,\infty)$ is a subalgebra of the Kac-Moody 
Lie algebra $L(\gA,\phi_1)$.

In order to formulate a general definition of automorphic Lie algebras we need 
to fix some notations. We will consider the groups 
$G$ whose elements are M\"obius (linear-fractional) transformations
\[  g_k(\lambda)=\frac{\alpha_k 
\lambda+\beta_k}{\gamma_k\lambda+\delta_k},\qquad 
\alpha_k\delta_k-\beta_k\gamma_k\ne 0,
\quad \alpha_k,\beta_k,\gamma_k,\delta_k\in\bbbc.
\]
of the extended complex plane $\cb=\bbbc\cup \{\infty\}$.
The group of all M\"obius transformations is called the M\"obius group which is 
isomorphic to $PSL(2,\bbbc)\simeq SL(2,\bbbc)/\pm I$,
where $SL(2,\bbbc)$ is a group of $2\times2$ matrices whose determinants are 
equal to $1$, and $I$ is the unit matrix. Indeed, if we 
associate a  matrix
\begin{equation}\label{mobmat}
S_k=\left(\begin{array}{cc}
\alpha_k &\beta_k\\
\gamma_k &\delta_k	
\end{array}\right)
\end{equation}
with the M\"obius transformation $g_k$, then the composition of transformations 
$g_p\cdot g_k$ corresponds to 
the product of the matrices $S_p S_k$. Matrices $S_k$ and $\theta S_k, \ 
\theta\ne 0,\ \theta\in\bbbc$ result in the same 
M\"obius transformation.

In this paper we are interested in finite subgroups of the M\"obius group. 
According to F.Klein \cite{klein}, 
all finite subgroups of $PSL(2,\bbbc)$ are in the following list:
\begin{enumerate}
\item the additive group of integers modulo $N$, $\mathbb{Z}/N\bbbz$
\item the symmetry group of the dihedron with $N$ vertices,
$\mathbb{D}_N$
\item the symmetry group of the tetrahedron, $\mathbb{T}$
\item the symmetry group of the octahedron, $\mathbb{O}$
\item the symmetry group of the icosahedron, $\mathbb{I}$
\end{enumerate}

In what follows we assume that $G$ is a finite group of M\"obius 
transformations. For any 
$\gamma_0\in\cb$ we denote the 
\textit{orbit} $G(\gamma_0)=\{g(\gamma_0) \,|\, g\in G\}$ and the 
\textit{isotropy subgroup}
$G_{\gamma_0} = \{g\in G \,|\, g(\gamma_0)=\gamma_0\}$. If the group 
$G_{\gamma_0}$ is nontrivial, i.e.  $|G_{\gamma_0}|>1$, 
then the point $\gamma_0$ is called a fixed point of the group $G$ of order  
$|G_{\gamma_0}|$. Points which are not fixed are 
called generic. Obviously, the number of points 
$|G(\gamma_0)|=|G|/|G_{\gamma_0}|$. If $\gamma_0$ is a fixed point of order 
$n$, 
then the corresponding orbit is called a degenerate orbit of degree $n$. We 
call orbits corresponding to generic points generic. 
Two  points $\gamma_0,\gamma_1\in\cb$ are said to be equivalent  
$\gamma_0\sim\gamma_1$ if they belong to the same orbit (for non equivalent 
points $\gamma_0,\gamma_1\in\cb$ we shall use the notation 
$\gamma_0\not\sim\gamma_1$). 

M\"obius transformations induce automorphisms of the field of rational 
functions $\bbbc(\lambda)$ defined as
\begin{equation}\label{sigmaf}
  g: \ f(\lambda)\rightarrow f(g^{-1}(\lambda)),\qquad f(\lambda)\in 
\bbbc(\lambda).
\end{equation}
If $G$ is a finite group of M\"obius transformations, then there exists a 
subfield $\bbbc^G(\lambda)$ of $G$-invariant rational functions 
\[ \bbbc^G(\lambda)=\{f\in\bbbc(\lambda)\,|\, g(f)=f,\ \forall g\in G\}
\]
Non constant elements of $\bbbc^G(\lambda)$ are called rational automorphic 
functions of the group $G$. Moreover, there 
exists a {\em primitive} automorphic function $J\in \bbbc^G(\lambda)$, such 
that any rational automorphic function is a 
rational function of $J$, or $\bbbc^G(\lambda)=\bbbc(J)$ (see 
\cite{LM05}). The primitive
automorphic function $J$ is not uniquely defined - any non constant fractional 
linear function of $J$ is a primitive automorphic function.

For finite groups automorphic functions can be easily constructed using the 
group average
\[\left<f(\lambda)\right>_G = \frac{1}{|G|} \sum_{g\in G} f(g^{-1}(\lambda)).\]
If an automorphic function has a pole  (or a zero) in $\lambda$ at a point 
$\lambda=\mu$, then its order is divisible by  $|G_{\mu}|$.

Let $\gamma_0 \in 
\mathbb{C}$, then
\[J_{G}(\lambda,\gamma_0) =
\left<\frac{1}{   (\lambda-\gamma_0)^{|G_{\gamma_0}|}  }\right>_G\]
is a primitive automorphic function which has poles of multiplicity 
$|G_{\gamma_0}|$ at the points of the orbit
$G(\gamma_0)$. If $\gamma_0=\infty$, then $J_{G}(\lambda,\infty) =
\left<\lambda^{|G_{\infty}|} \right>_G$.        

Assuming $\gamma_0\not\sim\gamma_1$ we define  a primitive automorphic function
\begin{equation}\label{J}
J_{G}(\lambda,\gamma_0,\gamma_1)=J_{G}(\lambda,\gamma_0)-J_{G}(\gamma_1,
\gamma_0)
\end{equation}
 with poles of multiplicity $|G_{\gamma_0}|$ at points of the orbit
$G(\gamma_0)$, zeros of multiplicity $|G_{\gamma_1}|$ at points of 
$G(\gamma_1)$ and no other poles or zeros.  Any primitive automorphic function 
with a pole at $\gamma_0$ and a zero at $\gamma_1$ is proportional to 
$J_{G}(\lambda,\gamma_0,\gamma_1)$. The following Lemma summarises some useful 
properties of the function $J_{G}(\lambda,\gamma_0,\gamma_1)$.

\begin{Lem}
 Let $G$ be a finite group, $\beta\not\sim\alpha $, $\beta\not\sim\gamma$ and 
$\beta\not\sim\delta$, then
\begin{eqnarray}\label{J+J}
&& J_{G}(\alpha,\beta,\gamma)+J_{G}(\gamma,\beta,\alpha)=0
\\ \label{J_zero}
&& 
J_{G}(\alpha,\beta,\gamma)-J_{G}(\delta,\beta,\gamma)=J_{G}(\alpha,\beta,\delta)
\\ && \label{JxJ0}
  J_{G}(\alpha,\beta,\gamma)J_{G}(\alpha,\gamma,\beta)=
C(\beta,\gamma)
\\ && \label{JxJ1}
  J_{G}(\alpha,\beta,\gamma)J_{G}(\alpha,\gamma,\delta)=
J_{G}(\alpha,\beta,\delta)J_{G}(\beta,\gamma,\delta)
\end{eqnarray}
where $C(\beta,\gamma)=C(\gamma,\beta)\ne 0$ and  $C(\beta,\gamma)$ does not 
depend on $\alpha$.
\end{Lem}
{\bf Proof.} Identity (\ref{J+J}) follows from (\ref{J_zero}) if we take 
$\gamma\sim\alpha$, and (\ref{J_zero}) immediately follows from (\ref{J}). The 
left hand side of (\ref{JxJ0}) is a product of two rational functions of 
$\alpha$.   Poles of $J_{G}(\alpha,\beta,\gamma)$ are all at $\alpha\sim\beta$ 
and are canceled by the corresponding zeros of $J_{G}(\alpha,\gamma,\beta)$. 
Similarly poles of $J_{G}(\alpha,\gamma,\beta)$ are all at $\alpha\sim\gamma$ 
and they are canceled by the corresponding zeros of 
$J_{G}(\alpha,\beta,\gamma)$. Thus the product is a rational automorphic 
function of $\alpha$ which does not have any poles. Therefore it is a 
constant function of $\alpha$. The property $C(\beta,\gamma)=C(\gamma,\beta)$ 
is obvious from the symmetry. Identity (\ref{JxJ1}) follows from 
(\ref{J+J})-(\ref{JxJ0}):
\[ 
J_{G}(\alpha,\beta,\gamma)J_{G}(\alpha,\gamma,\delta)=J_{G}(\alpha,\beta,
\gamma)(J_{G}(\alpha,\gamma,\beta)-J_{G}(\delta,
\gamma,\beta))=
\]
\[
 C(\beta,\gamma)-(J_{G}(\alpha,\beta,\delta)+J_{G}(\delta,\beta,\gamma)) 
J_G(\delta,\gamma,\beta)= 
C(\beta,\gamma)+J_{G}(\alpha,\beta,\delta)J_G(\beta,\gamma,\delta)-C(\beta,
\gamma)\, .\ \blacksquare
\]

If the group $G$ is clearly specified or the result is general and does not 
depend on the choice of a finite group $G$, we shall use a simplified notation 
by omitting the subscript $G$ in $J_G(\alpha,\beta,\gamma)$.

Let $\Gamma=G(\gamma_0)$ be an orbit of a finite subgroup $G\subset 
PSL(2,\bbbc)$ and $\cR_{\lambda}(\Gamma)$ the corresponding ring of 
rational functions with poles at $\Gamma$ only. Then  $G$ is a group of 
automorphisms of $\cR_{\lambda}(\Gamma)$. Indeed, the transformations 
(\ref{sigmaf}) map $\cR_{\lambda}(\Gamma)\rightarrow\cR_{\lambda}(\Gamma)$ and 
respect the ring structure. The $G$-invariant subring 
\[ \cR^G_{\lambda}(\Gamma)=\{a\in \cR_{\lambda}(\Gamma)\,|\, g(a)=a, \forall 
g\in G\}\]
is the ring of polynomials $\bbbc[J]$ of a primitive automorphic function 
$J=J_{G}(\lambda,\gamma_0)$.

Let us have a simple Lie algebra $\gA$, the corresponding generalised loop 
algebra $\gA_{\lambda}(\Gamma)$  (\ref{gAG}) and a homomorphism 
$\Psi: \,G\rightarrow \mbox{Aut}\,\gA_{\lambda}(\Gamma)$. We denote by $\cG$ the image $\Psi (G)$ 
in $\mbox{Aut}\,\gA_{\lambda}(\Gamma)$ and 
call it the {\em reduction group}. An element $\Phi_k\in\cG$  can be viewed as a 
pair 
$\Phi_k=(g_k,\phi_k)$ consisting of a M\"obius transformation $g_k$ and an automorphism $\phi_k\in\mbox{Aut}\, \gA$, which could depend on $\lambda$. The action of 
$\Phi_k$ on the elements of $\gA_{\lambda}(\Gamma)$ is similar to 
(\ref{autPhi2}):
\[
\Phi_k(a(\lambda))=\phi_k(a(g_k^{-1}(\lambda)))\, ,\qquad 
a(\lambda)\in\gA_{\lambda} (\Gamma).
\]

When the elements $\phi_k\in\mbox{Aut}\, \gA$ do not depend on $\lambda$ we can (without loss of generality - see \cite{LM05}) construct the reduction group as follows. Suppose we have a $\lambda$-independent homomorphism 
$\psi\, :\, G\rightarrow  \mbox{Aut}\,\gA$.  
For every element $g_k\in G$ we 
define $\Phi_k=(g_k,\psi_{g_k})$ and the reduction group 
$\cG(G,\psi)=\{\Phi=(g,\psi_g)\,|\, 
g\in G\}$, which is a subgroup of the direct product $G \times\mbox{Aut}\,\gA$ and is isomorphic to $G$.

The {\em automorphic Lie algebra} corresponding to the group $\cG$ and the 
orbit $\Gamma$ is the $\cG$-invariant subalgebra 
$\gA^{\cG}_{\lambda} (\Gamma)=\{a\in \gA_{\lambda}(\Gamma)\,|\, \Phi(a)=a, 
\forall \Phi\in \cG\}$. 

More generally, if ${\bf \Gamma}$ is a union of $M$ orbits ${\bf 
\Gamma}=\bigcup_{k=1}^M\Gamma_k$ of $G$, then
$G$ is still a group of automorphisms of the corresponding ring 
$\cR_{\lambda}({\bf \Gamma})$ and $\cG=\Psi(G)$ is a group of 
automorphisms of the Lie algebra $\gA_{\lambda} ({\bf \Gamma})$. We shall call
the $\cG$--invariant subalgebra $\gA^{\cG}_{\lambda} ({\bf \Gamma})$ the {\em
automorphic Lie algebra corresponding to the reduction group $\cG$ and orbits
$\Gamma_1, \ldots ,\Gamma_M$}. It is obvious that the subalgebras
$\gA^{\cG}_{\lambda} (\Gamma_k)\subset \gA^{\cG}_{\lambda} ({\bf \Gamma}), \
k=1,\ldots, M$  form a coverage of $\gA^{\cG}_{\lambda} ({\bf \Gamma})$:
\[ 
 \gA^{\cG}_{\lambda} ({\bf \Gamma})=\bigcup_{k=1}^M \gA^{\cG}_{\lambda} 
(\Gamma_k)\, ,\qquad 
 \gA^{\cG}_{\lambda} (\Gamma_k)\bigcap  \gA^{\cG}_{\lambda} 
(\Gamma_n)=\gA^{\cG}\, ,\ k\ne n .
\]

In this sense, the Kac-Moody Lie algebra $L(\gA,\phi_1)$ is  the automorphic 
Lie algebra corresponding to the cyclic group 
$\bbbz/n\bbbz$ and  ${\bf \Gamma}=\Gamma_1\cup\Gamma_2$, where  
$\Gamma_1=\{\infty\}$ and $\Gamma_2=\{0\}$ (of the M\"obius transformation 
$g_1(\lambda)=\omega\lambda$). 
Its subalgebra $L_+(\gA,\phi_1)$ is a $\bbbz/n\bbbz$--automorphic Lie algebra 
corresponding to one orbit $\Gamma_1$. Similarly,  
$L_-(\gA,\phi_1)$ corresponds to the orbit $\Gamma_2$. The algebra 
$\gA^{\cG}_{\lambda}(0,\infty)$ (\ref{sl2d2}) is the 
automorphic Lie algebra corresponding to the group $\cG\simeq\bbbd_n$ and a 
single degenerate  orbit $\Gamma=\{0,\infty\}$ of degree $n$.

There is a natural projection $\cP_\cG$ of the linear space $\gA_{\lambda} 
({\bf \Gamma})$ onto $\gA^{\cG}_{\lambda} ({\bf \Gamma})$ 
given by the group average. For $a\in\gA_{\lambda} ({\bf \Gamma})$ we define 
$\cP_\cG (a)\in \gA^{\cG}_{\lambda} ({\bf \Gamma})$ as
\begin{equation}\label{gav}
 \cP_\cG(a)=\langle a\rangle_{\cG}=\frac{1}{|\cG|}\sum_{\Phi\in\cG}\Phi(a)\, .
\end{equation}
Obviously $\cP_\cG^2=\cP_{\cG}$. The projection $\cP_\cG:\gA_{\lambda} ({\bf 
\Gamma})\rightarrow\gA^{\cG}_{\lambda} ({\bf \Gamma})$ 
is a surjective linear map, but it is {\em not} a Lie algebra homomorphism.

\section{Automorphic Lie algebras in the case $\gA=sl(2,\bbbc)$}

As above, $G$ denotes a finite group of M\"obius transformations. 
In the case $\gA=sl(2,\bbbc)$ it is well known (\cite{jacobson}, \cite{kac94}, \cite{carter}) 
that all automorphisms are 
inner and can be represented in the form
$a\rightarrow UaU^{-1}$ where $U\in GL(2,\bbbc)$. We shall denote such an 
automorphism as $\phi_U$, where $\phi_U(a)=UaU^{-1}$. Thus 
$\mbox{Aut}\,\gA\simeq PSL(2,\bbbc)$. Let us take any {\sl 
injective}\,\footnote{If the homomorphism $\rho$ is not injective and therefore 
the corresponding projective representation is not faithful, then its kernel 
$\mbox{ker}\, \rho\subset G$ is a normal subgroup in $G$ and the problem can be 
effectively reduced to the quotient group $G/\mbox{ker}\, \rho$ (see 
\cite{LM05}). } homomorphism $\rho:G\rightarrow PSL(2,\bbbc)$ (which can 
be regarded as a faithful projective representation $\rho:G\mapsto 
\mbox{End}\bbbc^2$) and define a homomorphism
$\psi_{\rho}:G\rightarrow \mbox{Aut}(sl(2,\bbbc))$ by its action on the 
M\"obius transformations  $g\in G:\ \psi_{\rho}(g)=\phi_{\rho(g)}$. Thus with 
any M\"obius group $G$ and a projective representation $\rho$ we associate a 
reduction group $\cG=\{ \Phi_k=(g, \phi_{\rho(g)})\,|\,g\in G\}$.

In the case $\gA=sl(2,\bbbc)$ there is
a natural homomorphism $\psi:G\rightarrow \mbox{Aut}(sl(2,\bbbc))$, namely 
$ \psi(g_k)=\phi_{S_k},$
where $g_k\in G$ and $S_k$ is the matrix (\ref{mobmat}) associated to the 
M\"obius transformation $g_k$. We call the reduction group 
$\cG=\{(g_k,\phi_{S_k})\,|\,g_k\in G\}$ the natural reduction group.

\subsection{The case $G=\bbbd_2$ and $\gA=sl(2,\bbbc)$} \label{sec3.1}

Without loss of generality we can represent the generators 
of the group $G$ by the M\"obius transformations
\[ g_1(\lambda)=-\lambda,\qquad g_2(\lambda)=\lambda^{-1}.\]

Thus the (natural) reduction group  $\cG\sim \bbbd_2$ is generated by the 
transformations
\begin{equation}
 \label{redgrD2}
 \Phi_1(a(\lambda))=\bbs_3 a(-\lambda)\bbs_3 ,\qquad  
\Phi_{2}(a(\lambda))=\bbs_1 a(\lambda^{-1})\bbs_1 .
\end{equation}
Here we use the notation
\[ {\bf s}_0= \left(\begin{array}{cc}
1&0\\0&1
\end{array}\right),\quad {\bf s}_1= \left(\begin{array}{cc}
0&1\\1&0
\end{array}\right),\quad {\bf s}_2= \left(\begin{array}{cc}
0&-1\\1&0
\end{array}\right),\quad {\bf s}_3= \left(\begin{array}{cc}
1&0\\0&-1
\end{array}\right) \, .
\]
The set $\{\bbs_0,\bbs_1,\bbs_2,\bbs_3\}$ is a two-dimensional irreducible 
projective representation of the group $\bbbd_2$. Since the group $\bbbd_2$ is 
commutative, all its linear irreducible representations are one-dimensional.

In order to construct a corresponding automorphic Lie algebra we need to choose 
an orbit $\Gamma$ of the group $G$ or a finite union of orbits. 
There are three degenerate orbits $\Gamma_0,\Gamma_1$ and $\Gamma_i$ of degree 2
\[ \Gamma_0=\{0,\infty\},\quad \Gamma_1=\{\pm 1\},\quad \Gamma_i=\{\pm i\}\] 
and a generic orbit \[\Gamma_{\mu}=\{\pm \mu,\pm\mu^{-1}\},\qquad 
\mu\not\in\{0,\infty,\pm 1,\pm i\}.\]

Elements of a basis of the automorphic Lie algebra 
$\gA_{\lambda}^\cG(\Gamma_0)$ can be constructed using the group average 
(\ref{gav}). We define ${\bf e}^1= 2\langle\lambda {\bf e}\rangle_{\cG},\ {\bf 
f}^1= 2\langle\lambda {\bf f}\rangle_{\cG},\ {\bf h}^2= 2 \langle\lambda^2 {\bf 
h}\rangle_{\cG}$. Evaluating the group average we get:
\begin{equation}
 {\bf e}^1=\left(\begin{array}{cc}
0&\lambda\\ \lambda^{-1}&0
\end{array}\right),\quad {\bf f}^1=\left(\begin{array}{cc}
0&\lambda^{-1}\\ \lambda&0
\end{array}\right),\quad  {\bf 
h}^2=(\lambda^2-\lambda^{-2})\left(\begin{array}{cc}
1&0\\0&-1
\end{array}\right) \, .
\label{bas1}
\end{equation}

Their  commutators   are
\begin{equation}\label{com1}
[ \bbe^1,\bbf^1 ] =  \bbh^{2}\, ,\quad
[ \bbh^2,\bbe^1 ] =  2(\lambda^2+\lambda^{-2})\bbe^{1}-4\bbf^{1}\, ,\quad
[ \bbh^2,\bbf^1 ] =  -2(\lambda^2+\lambda^{-2})\bbf^{1}+4\bbe^{1}\, .
\end{equation}
For $\nu\not\sim 0$ we define a primitive automorphic function 
$J_G(\lambda,0,\nu)=J_G(\lambda,0)-J_G(\nu,0)$ where $J_G (\lambda,0)=\langle 
\lambda^{-2}\rangle _{G}=\frac{1}{2}(\lambda^2+\lambda^{-2})$ (see (\ref{J})). 
The set 
\begin{equation}\label{bas}
 B=\bigcup_{n\in\bbbn}B_n\, ,\qquad B_n=\{ {\bf e}^{2n-1}=J^{n-1}{\bf 
e}^1,\quad {\bf f}^{2n-1}=J^{n-1} {\bf f}^1,\quad
{\bf h}^{2n}=J^{n-1} {\bf h}^2\},
\end{equation}
where $J=2J_G(\lambda,0,\nu)$
is a basis of $\gA_{\lambda}^\cG(\Gamma_0)$ (see \cite{LM05}).

It follows from the commutation relations (\ref{com1}) that
\begin{equation}\label{comm1}
\begin{array}{l}

[\bbe^n,\bbf^m]=\phantom{-}\bbh^{n+m}\, ,\\

[\bbh^k,\bbe^n]=\phantom{-}2\bbe^{n+k}-4\bbf^{n+k-2}+4J_G(\nu,0)\bbe^{n+k-2}\, 
,\\

[\bbh^k,\bbf^n]=-2\bbf^{n+k}+4\bbe^{n+k-2}-4J_G(\nu,0)\bbf^{n+k-2}\, ,
\end{array}
\end{equation}
where $n,m\in 2\bbbn-1$ and $ k\in 2\bbbn$. Thus
 the algebra $\gA_{\lambda}^\cG(\Gamma_0)$ is almost graded
\[ \gA_{\lambda}^\cG(\Gamma_0)=\bigoplus_{k=1}^{\infty}\cB^k,\quad 
[\cB^p,\cB^q]\subset\cB^{p+q}\bigoplus\cB^{p+q-1}\]
where homogeneous subspaces are $\cB^p=\mbox{Span}_{\bbbc}(B_p)$.    If we set 
$\nu=\exp\frac{i\pi}{4}$ then $J_G(\nu,0)=0$ and the commutation relations 
(\ref{comm1}) take a rather  simple form. A choice  of the point $\nu$ (which 
controls zeros of the automorphic function $J_G(\lambda,0,\nu)$) corresponds to 
a  choice of the basis in $\gA_{\lambda}^\cG(\Gamma_0)$. The grading structure 
depends on the choice of $\nu$ (see \cite{LM05}). It follows from the 
commutation relations (\ref{comm1}) that the algebra $ 
\gA_{\lambda}^\cG(\Gamma_0)$ is generated by its first homogeneous space 
$\cB^1$ (and actually, in this particular case, by two elements $\bbe^1$ and 
$\bbf^1$). 

{\bf Remark:} Almost graded algebras  can be seen as  deformations of the 
corresponding graded algebras. For example the
Automorphic Lie algebra $\gA_{\lambda}^\cG(\Gamma_0)$ is a deformation of the 
graded algebra
$ L_{>0}(\gA,\phi_1)=\bigoplus_{k> 0}\gA_k\subset L_{+}(\gA,\phi_1)$ 
(\ref{kacA1}). Indeed, after the re-scaling (which is a graded isomorphism)
\[ 
 \hat{\bbe}^n=\epsilon^n \bbe^n,\quad \hat{\bbf}^n=\epsilon^n \bbf^n,\quad 
\hat{\bbh}^n=\epsilon^n \bbh^n
\]
 the commutation relations  (\ref{comm1}) take the form
\[
\begin{array}{l}

[\hat{\bbe}^n,\hat{\bbf}^m]=\phantom{-}\hat{\bbh}^{n+m}\, ,\\

[\hat{\bbh}^k,\hat{\bbe}^n]=\phantom{-}2\hat{\bbe}^{n+k}-4\epsilon^2\hat{\bbf}^{
n+k-2}+4\epsilon^2J_G(\nu,0)\hat{\bbe}^{n+k-2}\, ,\\

[\hat{\bbh}^k,\hat{\bbf}^n]=-2\hat{\bbf}^{n+k}+4\epsilon^2\hat{\bbe}^{n+k-2}
-4\epsilon^2 J_G(\nu,0)\hat{\bbf}^{n+k-2}\, .
\end{array}
\]
Setting (formally) $\epsilon=0$, we obtain the commutation relations for the 
algebra $ L_{>0}(\gA,\phi_1)$. 

Similarly one can construct a basis for the algebra 
$\gA_{\lambda}^\cG(\Gamma)$, for any orbit $\Gamma=G(\kappa)$ and compute the 
corresponding structure constants.
For example a basis for $\gA_{\lambda}^\cG(\Gamma_1)$ can be chosen as:
\[\{\hat{J}^{n-1} \cP_\cG\left(\frac{{\bf e}}{\lambda-1} \right),\ 
\hat{J}^{n-1} \cP_\cG\left(\frac{ {\bf f}}{(\lambda-1)^2}\right),\ 
\hat{J}^{n-1} \cP_\cG\left(\frac{{\bf h}}{\lambda-1}  \right)\,|\, 
n\in\bbbn\},\]
where $\hat{J}=J_G(\lambda,1,\nu),\ \nu\not\sim 1$.

There is, however, a more elegant way to give a description of the automorphic 
Lie algebra $\gA_{\lambda}^\cG(\Gamma)$ for any orbit $\Gamma=G(\kappa),\ 
\kappa\not\sim 0$.

\begin{Pro}\label{prop1}
 Let $\cG\simeq \bbbd_2$ be the reduction group generated by the automorphisms 
(\ref{redgrD2}), $\kappa\in\bbbc \setminus\{0,\infty\}$, 
$\Gamma_{\kappa}=G(\kappa)$  and $J_{\kappa}=J_G(\lambda,\kappa,0)$. Then 
\begin{itemize}
 \item[(i)]
the automorphic Lie algebra 
$\gA_{\lambda}^\cG(\Gamma_{\kappa})$ is generated by  
$$\bba_1=J_{\kappa}\bbe^1, \quad \bba_2=J_{\kappa}\bbf^1, \quad 
\bba_3=J_{\kappa}\bbh^2.$$
\item[(ii)] The commutation relations between the generators are
\begin{eqnarray}\label{commGamma12}
&&[\bba_1 ,\bba_2 ]=\phantom{-}J_{\kappa}\bba_3 ,\\
\label{commGamma31}&&
[\bba_3,\bba_1]=\phantom{-}2 (\kappa^2+\kappa^{-2})J_{\kappa}\bba_1 
-4J_{\kappa}\bba_2+4C(0,\kappa)\bba_1 ,\\
\label{commGamma32}&&
[\bba_3,\bba_2]=-2 (\kappa^2+\kappa^{-2})J_{\kappa}\bba_2 
+4J_{\kappa}\bba_1-4C(0,\kappa)\bba_2 .
\end{eqnarray}
\item[(iii)] The set 
\begin{equation}\label{basB}
  B=\bigcup_{n\in\bbbn}B_n\, ,\ B_n=\{ J_{\kappa}^{n-1}\bba_1,\ 
J_{\kappa}^{n-1}\bba_2,\ J_{\kappa}^{n-1}\bba_3\}
\end{equation}
is a basis of $\gA_{\lambda}^\cG(\Gamma_{\kappa})$.
\item[(iv)] The algebra $\gA_{\lambda}^\cG(\Gamma_{\kappa})$ is almost graded
\[ \gA_{\lambda}^\cG(\Gamma_{\kappa})=\bigoplus_{k=1}^{\infty}\cB^k,\quad 
[\cB^p,\cB^q]\subset\cB^{p+q}\bigoplus\cB^{p+q-1}\]
where $\cB^k=\mbox{\rm Span}_{\bbbc}(B_k)$.
\end{itemize}
\end{Pro}
{\bf Proof.} $(ii)$: The commutation relation (\ref{commGamma12}) immediately 
follows from (\ref{comm1}):
\[ 
[\bba_1 ,\bba_2 ]=J_{\kappa}^2[\bbe^1,\bbf^1]=J_{\kappa}^2 
\bbh^2=J_{\kappa}\bba_3,
\]
To show (\ref{commGamma31}),(\ref{commGamma32}) we use (\ref{comm1}) and 
(\ref{J}),(\ref{JxJ0}). To demonstrate (\ref{commGamma31}) we note:
\[
 [\bba_3,\bba_1]=J_{\kappa}^2[ \bbh^2,\bbe^1 ] =  
2(\lambda^2+\lambda^{-2})J_{\kappa}\bba_{1}-4J_{\kappa}\bba_{2}
\]
We recall that
$(\lambda^2+\lambda^{-2})J_{\kappa}=2J_G(\lambda,0)J_G(\lambda,\kappa,0)$. It
follows  from Lemma 1 (\ref{J}),(\ref{JxJ0}) that 
\[
 2J_G(\lambda,0)J_G(\lambda,\kappa,0)=2J_G(\kappa,0)J_G(\lambda,\kappa,0)+
2J_G(\lambda,0,\kappa)J_G(\lambda,\kappa,0)=(\kappa^2+\kappa^{-2})J_{\kappa}
+2C(0,\kappa)
\]
and thus $[\bba_3,\bba_1]=2 (\kappa^2+\kappa^{-2})J_{\kappa}\bba_1 
-4J_{\kappa}\bba_2+4C(0,\kappa)\bba_1$. The proof of (\ref{commGamma32}) is 
similar.

$(iii)$: The elements $J_{\kappa}^{n-1}\bba_i$ are $\cG$ invariant and have 
poles at points of $\Gamma_{\kappa}$ only, thus  
$J_{\kappa}^{n-1}\bba_i\in\gA_{\lambda}^\cG(\Gamma_{\kappa})$. It is easy to 
show that any element of $\gA_{\lambda}^\cG(\Gamma_{\kappa})$ can be 
represented as a finite linear combination of elements of $B$. For a generic 
point $\kappa$, the proof of the latter statement is given in \cite{LM05} 
(Proposition 3.1). In the case of degenerate orbits  $\Gamma_1,\Gamma_i$ the 
proof is similar (or can be deduced from the generic orbit case).  Thus $B$ is 
a basis of $\gA_{\lambda}^\cG(\Gamma_{\kappa})$.

$(i)$: It follows from $(ii)$ that all elements of $B$ can be generated by the 
set $B_1=\{\bba_1,\bba_2,\bba_3\}$.

$(iv)$: Since $B$ is a basis of $\gA_{\lambda}^\cG(\Gamma_{\kappa})$, we have 
$\gA_{\lambda}^\cG(\Gamma_{\kappa})=\bigoplus_{k=1}^{\infty}\cB^k$ where 
$\cB^k=\mbox{\rm Span}_{\bbbc}(B_n)$.  It follows from $(ii)$ that 
$[\cB^p,\cB^q]\subset\cB^{p+q}\bigoplus\cB^{p+q-1}$.\hfill $\blacksquare$

It follows from the above Proposition and the preceding discussion that for any orbit $\Gamma$ an 
almost graded basis of the  algebra $\gA_{\lambda}^\cG(\Gamma)$ can be 
characterised by a set of generators $\{\bba_1, \bba_2, \bba_3\}$ and a 
primitive automorphic function $J$ with poles at $\Gamma$ and thus it is 
convenient to introduce the notation
\[ \langle \bba_1, \bba_2, \bba_3\, ;\, J\rangle=\{J^{n-1}\bba_i\,|\, 
n\in\bbbn, i=1,2,3\}.\]
In this notation $B=\langle \bba_1, \bba_2, \bba_3\, ;\, J_{\kappa}\rangle$ (\ref{basB}). 

An almost graded $\bbbc$--algebra with the basis $\langle \bba_1, \bba_2, 
\bba_3\, ;\, J\rangle$ is infinite dimensional and can be viewed as a $\bbbc[J]$--Lie module with three generators. It can be completely 
characterised by a finite number of structure constants $C^1_{ijk},C^0_{ijk}\in 
\bbbc\, $:
\begin{equation}
 [\bba_i,\bba_j]=\sum_{k} C^1_{ijk}J\bba_k+C^0_{ijk}\bba_k \, 
.\label{struct}
\end{equation}
Two algebras are isomorphic  iff there exist bases such that the corresponding 
structure constants coincide.
\begin{Def}
Given two almost graded Lie algebras $\cA$ and $\cB$ with bases defined by 
$\langle\mathbf{a}_1,\ldots,\mathbf{a}_N\,;\,J_a\rangle$ and 
$\langle\mathbf{b}_1,\ldots,\mathbf{b}_N\,;\,J_b\rangle$ respectively, we say 
that the algebras are {\em graded~isomorphic} if there exists a linear 
transformation of the form
\begin{equation}
 \label{lintrans}
\hat{\mathbf{b}}_i=\sum_{k=1}^N W_{ik}\mathbf{b}_k,\quad  \hat{J}_b=\Delta J_b 
+ \delta,\qquad   \Delta,\delta, W_{ij}\in\bbbc, \quad \det W\ne 0,
\end{equation}
such that the structure constants of the algebras $\cA$ and $\cB$ in the bases 
$\langle\mathbf{a}_1,\ldots,\mathbf{a}_N\,;\,J_a\rangle$ and 
$\langle\hat{\mathbf{b}}_1,\ldots,\hat{\mathbf{b}}_N\,;\,\hat{J}_b\rangle$
coincide.
\end{Def}

Graded--isomorphic algebras are of course isomorphic. The advantage of the 
graded~isomorphism is that it can be effectively verified for infinite 
dimensional almost graded $\bbbc$--algebras. Suppose we are given two almost 
graded algebras, one with the basis $\langle \bba_1, \ldots, \bba_N\, ;\, 
J_{a}\rangle$, commutation relations (\ref{struct}) and thus with structure 
constants $C^1_{ijk},C^0_{ijk}$ and another one $\langle \bbb_1, \ldots, 
\bbb_N\, ;\, J_{b}\rangle$ with commutation relations 
\[ 
 [\bbb_i,\bbb_j]=\sum_{k=1}^N S^1_{ijk}J_{b}\bbb_k+S^0_{ijk}\bbb_k 
\]
and corresponding structure constants $S^1_{ijk}$ and $S^0_{ijk}$. Then  for 
the transformed elements $\hat{\bbb}_p=\sum_{i=1}^N W_{pi}\bbb_i$ and 
$J_b=\Delta^{-1}(\hat{J}_b-\delta)$ we have:
\[ [\hat{\bbb}_p,\hat{\bbb}_q]=\sum_{i,j,k,s=1}^N 
\frac{\hat{J}_{b}}{\Delta}W_{pi}W_{qj}S^1_{ijk}W^{-1}_{ks}\hat{\bbb}_s-\frac{
\delta}{\Delta}W_{pi}W_{qj}S^1_{ijk}W^{-1}_{ks}\hat{\bbb}_s+W_{pi}W_{qj}S^0_{ijk
}W^{-1}_{ks}\hat{\bbb}_s  \]
Equating the transformed structure constants with  $C^1_{ijk},C^0_{ijk}$ we 
obtain the following overdetermined  system of $N^2(N-1)$ polynomial equations 
\begin{equation}\label{P1P0}
 P^0_{pqk}=0,\quad P^1_{pqk}=0,\qquad p,q,k\in\{1,\ldots ,N\},\ \ p>q,
\end{equation}
where
\begin{eqnarray}\label{weq1}
 P^1_{pqk}&=&\sum_{i,j=1}^N W_{pi}W_{qj}S^1_{ijk}-\sum_{s=1}^N \Delta 
C^1_{pqs}W_{sk},\\
\label{weq2}
 P^0_{pqk}&=&\sum_{i,j=1}^N W_{pi}W_{qj}S^0_{ijk}-\sum_{s=1}^N 
(C^0_{pqs}+\delta  C^1_{pqs})W_{sk}
\end{eqnarray}
for $N^2+2$ unknowns $W_{ij},\Delta,\delta$.

\begin{Pro}\label{pro2} Let $\gA=sl(2,\bbbc)$ and $\cG\simeq \bbbd_2$ be the 
reduction group generated by the automorphisms (\ref{redgrD2}).
The automorphic Lie algebras $\gA_{\lambda}^\cG(\Gamma_0)$, 
$\gA_{\lambda}^\cG(\Gamma_1)$ and $\gA_{\lambda}^\cG(\Gamma_i)$, corresponding 
to degenerate orbits, are graded~isomorphic.
\end{Pro}
{\bf Proof:} The algebra $\gA_{\lambda}^\cG(\Gamma_0)$ is almost graded in the 
basis $\langle \bbe^1, \bbf^1, \bbh^2\, ;\, J_0\rangle$ with structure 
constants defined by (\ref{com1})
\begin{equation}\label{com1J}
[ \bbe^1,\bbf^1 ] =  \bbh^{2}\, ,\quad
[ \bbh^2,\bbe^1 ] =  4J_0\bbe^{1}-4\bbf^{1}\, ,\quad
[ \bbh^2,\bbf^1 ] =  -4J_0\bbf^{1}+4\bbe^{1}\, .
\end{equation} 
It follows from Proposition \ref{prop1} that the algebra 
 $\gA_{\lambda}^\cG(\Gamma_1)$  is almost graded in the basis $\langle \bba_1, 
\bba_2, \bba_3\, ;\, J_{1}\rangle$
with structure constants defined by  (\ref{commGamma12}),(\ref{commGamma31}) 
and (\ref{commGamma32}), ($\kappa=1,\ C(0,1)=1$):
\[
[\bba_1 ,\bba_2 ]=J_1\bba_3 ,\
[\bba_3,\bba_1]=4J_1\bba_1 -4J_1\bba_2+4\bba_1 ,\
[\bba_3,\bba_2]=-4J_1\bba_2 +4J_1\bba_1-4\bba_2 .
\]
It is easy to verify that the following invertible linear map  
$\gA_{\lambda}^\cG(\Gamma_1)\mapsto\gA_{\lambda}^\cG(\Gamma_0)$ 
\[\bbe_1 = \bba_1-\bba_2-\frac{1}{2}\bba_3,\ \ \bbf_1= 
-\bba_1+\bba_2-\frac{1}{2}\bba_3,\ \ \bbh_2=4\bba_1+4\bba_2,\ \ J_0= 8 J_1+2\,  
\]
is the graded~isomorphism.
If we denote by $\langle \hat{\bba}_1, \hat{\bba}_2, \hat{\bba}_3\, ;\, 
J_{i}\rangle$ the basis of $\gA_{\lambda}^\cG(\Gamma_i)$ with structure 
constants defined by  (\ref{commGamma12}),(\ref{commGamma31}) and 
(\ref{commGamma32}) and $\kappa=i,\ C(0,i)=1$, then the linear map 
$\gA_{\lambda}^\cG(\Gamma_1)\mapsto \gA_{\lambda}^\cG(\Gamma_i)$ given by
$ \hat{\bba}_1=-\bba_1, \ \hat{\bba}_2=\bba_2,\ \hat{\bba}_3=\bba_3, J_i=-J_1$ 
is a graded isomorphism.\hfill $\blacksquare$

\begin{Pro}\label{pro3}
 Let $\gA=sl(2,\bbbc)$ and  $\cG\simeq \bbbd_2$ be the reduction group 
generated by the automorphisms (\ref{redgrD2}).
The automorphic Lie algebras $\gA_{\lambda}^\cG(\Gamma_\mu)$, 
$\gA_{\lambda}^\cG(\Gamma_\nu)$ are graded~isomorphic if and only if 
\begin{equation}
 \label{munu}
\nu\in G(\mu)\cup G(i\mu)\cup G(\frac{\mu-1}{\mu +1})\cup G(i\frac{\mu-1}{\mu 
+1})\cup G(\frac{i\mu-1}{i\mu +1})\cup G(i\frac{i\mu-1}{i\mu +1})\, .
\end{equation}
\end{Pro}
In the proof of the proposition we shall use the following Lemma.

\begin{Lem}\label{lem2}
Consider two almost graded algebras $\cA$ and $\cB$ with bases
$\langle\mathbf{a}_1,\mathbf{a}_2,\mathbf{a}_3\,;\,J_a\rangle$ and
$\langle\mathbf{b}_1,\mathbf{b}_2,\mathbf{b}_3\,;\,J_b\rangle$ and the
commutation relations
\begin{equation}\label{alphabeta}
  \begin{array}{ll}
&[\bba_1 ,\bba_2 ]=J_a\bba_3,  \\
&[\bba_3,\bba_1]=4\alpha J_a\bba_1 -4J_a\bba_2+4\bba_1 ,\\
&[\bba_3,\bba_2]=-4\alpha J_a\bba_2 +4J_a\bba_1-4\bba_2 ,
\end{array}
\qquad
 \begin{array}{ll}
&[\bbb_1 ,\bbb_2 ]=J_b\bbb_3 ,\\
&[\bbb_3,\bbb_1]=4\beta J_b\bbb_1 -4J_b\bbb_2+4\bbb_1 ,\\
&[\bbb_3,\bbb_2]=-4\beta J_b\bbb_2 +4J_b\bbb_1-4\bbb_2 ,
 \end{array}
\end{equation}
respectively. If algebras $\cA$ and $\cB$ are graded~isomorphic, then 
\begin{equation}\label{alphabetaequation}
 ( \alpha^2-\beta^2) ((\alpha+3)^2-\beta^2 (\alpha-1)^2) ((\alpha-3)^2-\beta^2 
(\alpha+1)^2)=0.
\end{equation}
\end{Lem}
{\bf Proof:} Algebras $\cA$ and $\cB$ are graded~isomorphic and thus there 
exists a solution of equations (\ref{P1P0}) for unknowns $W_{ij},\Delta,\delta$ 
with the condition $\det W=\gamma\ne 0$. Obviously $P^0_{pqk},P^1_{pqk},\ 
p,q,s\in\{1,2,3\}$ (\ref{weq1}), (\ref{weq2})  are polynomials in $W_{ij}, 
\alpha,\beta,\Delta,\delta$. In the polynomial  ring $\bbbc 
[W_{ij},\alpha,\beta,\gamma,\Delta,\delta]$ we consider the ideal $\cJ=\langle 
P^0_{pqk},P^1_{pqk}, \det W-\gamma\rangle$ generated by all polynomials 
$P^0_{pqk},P^1_{pqk},\ p,q,s\in\{1,2,3\}$ and the polynomial $\det W-\gamma$.
It can be duly shown that the polynomial
\begin{equation}\label{pi}
 \pi=\gamma ( \alpha^2-\beta^2) ((\alpha+3)^2-\beta^2 (\alpha-1)^2) 
((\alpha-3)^2-\beta^2 (\alpha+1)^2)\in\cJ
\end{equation}
belongs to the ideal $\cJ$. Thus $\pi=0$ for every solution of the system 
(\ref{P1P0}) and the equation $\det W=\gamma$. Since $\gamma\ne 0$ we get 
(\ref{alphabetaequation}).
\hfill $\blacksquare$

 {\bf Proof of Proposition \ref{pro3}:} First we show that the isomorphism 
$\gA_{\lambda}^\cG(\Gamma_\mu)\simeq\gA_{\lambda}^\cG(\Gamma_\nu)$ follows from
(\ref{munu}). It is sufficient to show that 
$\gA_{\lambda}^\cG(\Gamma_\mu)\simeq\gA_{\lambda}^\cG(\Gamma_\nu)$ for
 $\nu=i\mu$ and $\nu=\frac{\mu-1}{\mu +1}$. Indeed,  if $\nu\in G(\mu)$ then
$\Gamma_\nu=\Gamma_\mu$ and the algebras coincide, while the remaining cases can
be reduced to the above two cases by compositions. In the case of a degenerate
orbit $\Gamma_\mu$ condition (\ref{munu}) means that $\nu\in \Gamma_0\bigcup 
\Gamma_1 \bigcup \Gamma_i$ and the statement follows from Proposition 
\ref{pro2}. Thus we
shall assume that point $\mu$ is generic.   Since $C(0,\mu)\ne 
0$, by
the re-scaling $\bba_i\mapsto C(0,\mu)\bba_i,\ J \mapsto C(0,\mu)J_a$ we can
reduce the commutation relations  (\ref{commGamma12}),(\ref{commGamma31}) and
(\ref{commGamma32}) for the algebra $\gA_{\lambda}^\cG(\Gamma_\mu)$ to
(\ref{alphabeta}) for the algebra $\cA$ with
$\alpha=\frac{1}{2}(\mu^2+\mu^{-2})$. For a generic point $\mu$ we have
$\alpha\ne \pm 1$.
Similarly for the algebra $\gA_{\lambda}^\cG(\Gamma_\nu)$ we get structure 
constants for the algebra $\cB$  (\ref{alphabeta}) with
 $\beta=\frac{1}{2}(\nu^2+\nu^{-2})$. 

 In the case $\nu=i\mu$ we have $\beta=-\alpha$ and it is easy to verify that
the linear map
$\gA_{\lambda}^\cG(\Gamma_{i\mu})\mapsto\gA_{\lambda}^\cG(\Gamma_{\mu})$ of  the
form (\ref{lintrans})
\[
 \bba_1=-\bbb_1, \quad \bba_2=\bbb_2,\quad \bba_3=\bbb_3,\quad J_a=-J_b
\]
is the algebra homomorphism.
Similarly, 
 \[\bba_1 =
\frac{1}{1-\alpha}(\hat{\bba}_1-\hat{\bba}_2-\frac{1}{2}\hat{\bba}_3),\ \
\bba_2 = 
\frac{1}{1-\alpha}(-\hat{\bba}_1+\hat{\bba}_2-\frac{1}{2}\hat{\bba}_3),\ \
\bba_2 = \frac{1}{1-\alpha}(4\hat{\bba}_1+4\hat{\bba}_2) \]
 and $J_a = -\frac{4}{(\alpha-1)^2} J_{\hat{a}}-\frac{1}{\alpha-1}$ maps
the basis of $\gA_{\lambda}^\cG(\Gamma_{\mu})$ into the basis of
$\gA_{\lambda}^\cG(\Gamma_{\nu})$ with $\nu=\frac{\mu-1}{\mu +1}$.

 The necessity follows from the statement (\ref{alphabetaequation}) of Lemma \ref{lem2}. If the algebras $\cA$ and $\cB$ are graded~ isomorphic, then
$\beta\in\{\pm\alpha,\pm\frac{\alpha+3}{\alpha-1},\pm\frac{\alpha-3}{\alpha+1}\}
$. The case $\beta=\alpha$ corresponds to $\nu\in G(\mu)$, the cases 
$\beta=-\alpha,\frac{\alpha+3}{\alpha-1}, 
-\frac{\alpha+3}{\alpha-1},\frac{\alpha-3}{\alpha+1}$ and 
$-\frac{\alpha-3}{\alpha+1}$ correspond to $\nu\in G(i\mu), G(\frac{\mu-1}{\mu 
+1}), G(i\frac{\mu-1}{\mu +1}), G(\frac{i\mu-1}{i\mu +1})$ and 
$\nu\in G(i\frac{i\mu-1}{i\mu +1})$ respectively.
\hfill $\blacksquare$

The group of automorphisms $\mbox{Aut\, } \cG$ of the reduction group 
$\cG\simeq\bbbd_2$ (\ref{redgrD2}) is isomorphic to the dihedral group 
$\bbbd_3$, and it has six elements. Elements of $\mbox{Aut\, } \cG$ act by 
permutations on the set of the orbits $G(\mu), G(i\mu), G(\frac{\mu-1}{\mu 
+1}), G(i\frac{\mu-1}{\mu +1}), G(\frac{i\mu-1}{i\mu +1}), 
G(i\frac{i\mu-1}{i\mu +1})$. That explains the number of solutions to the 
equation (\ref{pi}).

In this Section we have shown that for $\gA=sl(2,\bbbc)$ and $\cG\simeq 
\bbbd_2$ (\ref{redgrD2}) there are two essentially different types of 
automorphic Lie algebra. The first one corresponds to  degenerate orbits of the 
M\"obius group, and the algebras corresponding to different degenerate orbits 
are all isomorphic. The second type is the automorphic Lie algebras  
$\gA_{\lambda}^\cG(\Gamma_\mu)$ corresponding to generic orbits of the M\"obius 
group $G$.  If $\mu$ and $\nu$ are two generic points, then the corresponding 
automorphic Lie algebras are graded~isomorphic if and only if the condition 
(\ref{munu}) is satisfied.
\begin{Pro}\label{pro4} Let $\gA=sl(2,\bbbc)$ and  $\cG\simeq \bbbd_2$ be the 
reduction group generated by the automorphisms (\ref{redgrD2}).
 Automorphic Lie algebras corresponding to generic and degenerate orbits are 
not isomorphic.
\end{Pro}
{\bf Proof:} 
 Assuming $\mu$ to be a generic point, it follows from 
(\ref{commGamma12}),(\ref{commGamma31}),(\ref{commGamma32}) and (\ref{com1})  
that 
$$\mbox{dim}\,  
(\gA_{\lambda}^\cG(\Gamma_\mu)\diagup\langle[\gA_{\lambda}^\cG(\Gamma_\mu),\gA_{
\lambda}^\cG(\Gamma_\mu)]\rangle)=3\,,\qquad \mbox{dim}\,(  
\gA_{\lambda}^\cG(\Gamma_0)\diagup\langle[\gA_{\lambda}^\cG(\Gamma_0),\gA_{
\lambda}^\cG(\Gamma_0)]\rangle)=2.$$
where 
$\langle[\gA_{\lambda}^\cG(\Gamma_\nu),\gA_{\lambda}^\cG(\Gamma_\nu)]\rangle$ 
denotes the ideal generated by the commutator of the algebra 
$\gA_{\lambda}^\cG(\Gamma_\nu)$ with itself.
\hfill  $\blacksquare$ 

\subsection{ Automorphic Lie algebras corresponding to finite reduction groups, 
$\gA=sl(2,\bbbc)$}

 In this section we consider automorphic Lie algebras
$\gA_{\lambda}^\cG(\Gamma)$ where $\gA=sl(2,\bbbc)$, $\Gamma$ is an orbit (with
respect to the M\"obius group associated with a finite reduction group $\cG$) of
a point, which may be either degenerate or generic. With every finite M\"obius
group and every  2-dimensional faithful projective representation of the group
we can associate a reduction group. Thus one could expect that the number of
automorphic Lie algebras is rather big. However, it turns out that there exist
many graded~isomorphisms between these   automorphic Lie
algebras.

\begin{Pro}\label{zngr}
Automorphic Lie algebras corresponding to groups $\mathbb{Z}_N$, $N\geq 2$ 
and the degenerate orbit $\Gamma=\{\infty\}$ are isomorphic.
\end{Pro}
{\bf Proof.} 
For a given $N\geq 2$ we choose the following generators:
\[\mathbf{a}_1=\lambda\mathbf{e},\quad\mathbf{a}_2=\lambda^{N-1}\mathbf{f},
\quad\mathbf{a}_3=\mathbf{h},\quad J=\lambda^N\]
then the commutation relations are
\[[\mathbf{a}_1,\mathbf{a}_2]=J\mathbf{a}_3,\qquad
[\mathbf{a}_3,\mathbf{a}_1]=2\mathbf{a}_1
,\qquad [\mathbf{a}_3,\mathbf{a}_2]=-2\mathbf{a}_2\]
and they do not depend on the choice of $N$.
These are the commutation relations of the Kac-Moody
subalgebra $L_+(\gA,\phi),\ \phi^2=id$.\hfill $\blacksquare$

Furthermore, automorphic Lie algebras corresponding to the dihedral groups
$\bbbd_N$ for any $N\ge 2$ and any irreducible projective representations are
graded~isomorphic to the cases considered in  Section \ref{sec3.1} for the
group
$\bbbd_2$ - with the distinction remaining between when we take $\Gamma$
degenerate or generic (for generic orbits this has been shown in
\cite{LM05}). 

There also exist graded~isomorphisms between all algebras associated with
non-cyclic reduction groups and degenerate orbits.

\begin{The}
Let $\gA=sl(2,\bbbc)$, $\cG$ be any finite non-cyclic reduction group and
$\Gamma$ be a degenerate orbit of the corresponding M\"obius group. Then the
automorphic Lie algebra $\gA_{\lambda}^\cG(\Gamma)$ is graded~isomorphic to the
algebra with $\cG\simeq\bbbd_2$ and the degenerate orbit $\Gamma=\{0,\infty\}$.
\end{The}

{\bf Sketch of the proof.} Our proof is elementary,  but long\footnote{The
conjecture that the algebras mentioned in the Theorem are isomorphic had been
formulated by one of the authors (AVM) in 2008. When our proof of the Theorem
was completed and announced at a number of seminars and the conference `Symmetry
in Nonlinear Mathematical Physics - 2009'', Kiev, we were informed that a
short and elegant proof of the conjecture has been done by S.Lombardo and
J.Sanders (now published in \cite{ls10}). Their proof is based on the
classical theory of invariants. In \cite{ls10} the authors also introduced a
canonical basis for automorphic Lie algebras, which is analogous to the
Cartan-Weyl basis.}. We consider all finite non-cyclic M\"obius groups, namely
the groups $\bbbd_N,\bbbt,\bbbo$ and $\bbbi$. We take (in turn) each one of the
groups. We take (in turn) each one of the three possible degenerate orbits
(these orbits are listed in \cite{LM05}, Appendix A). We consider (in
turn) all faithful $2$-dimensional projective representations of the chosen
group and construct the corresponding reduction groups. Taking each one of the
reduction groups and each orbit we evaluate the reduction group average to find
a basis of the associated automorphic Lie algebra $\gA_{\lambda}^\cG(\Gamma)$
and compute the corresponding structure constants. Finally we  find a linear
transformation (\ref{lintrans}) which transforms the structure constants
obtained to the structure constants (\ref{com1}) of the algebra with
$\cG\simeq\bbbd_2$ and the degenerate orbit $\Gamma=\{0,\infty\}$. 

For example, let us take the icosahedral group $G=\bbbi$. As a M\"obius group 
it can be generated by two linear-fractional transformations
\[ g_1(\lambda)=\varepsilon\lambda,\qquad 
g_2(\lambda)=\frac{(\varepsilon^2+\varepsilon^3)\lambda+1}{
\lambda-\varepsilon^2-\varepsilon^3} \, ,\quad \varepsilon=\exp\left(\frac{2 
\pi i}{5}\right).\]
The group $\bbbi$ has two 2-dimensional irreducible projective representations. 
Let us take the natural representation generated by 
\[ U_1=\left(\begin{array}{cc}
\varepsilon&0\\0&1
\end{array}\right)\,\qquad
U_2=\left(\begin{array}{cc}
\varepsilon^2+\varepsilon^3& 1\\1&-\varepsilon^2-\varepsilon^3
\end{array}\right).
\]
Thus the reduction group $\cG$ is generated by two automorphisms
\[ \Phi_1(a(\lambda))=U_1a(g_1^{-1}(\lambda))U_1^{-1},\qquad  
\Phi_2(a(\lambda))=U_2a(g_2^{-1}(\lambda))U_2^{-1}.\]
Let us choose the degenerate orbit of order $5$
\[ \Gamma=\{ 0,\infty, \varepsilon^{k+1}+\varepsilon^{k-1}, 
\varepsilon^{k+2}+\varepsilon^{k-2}\, |\, k=0,1,2,3,4\} \, .\]
The corresponding automorphic Lie algebra $\gA_{\lambda}^\cG(\Gamma)$ has 
generators

\[\mathbf{b}_1=\left<\lambda\mathbf{e}\right>_{\mathbb{I}},\quad 
\mathbf{b}_2=\left<\lambda^4\mathbf{f}\right>_{\mathbb{I}},\quad 
\mathbf{b}_3=\left<\lambda^5\mathbf{h}\right>_{\mathbb{I}},\quad 
J_\bbbi=\left<\lambda^5\right>_{\mathbb{I}}\]
with the following commutation relations
\[ 
[\mathbf{b}_1,\mathbf{b}_2]=\frac{5}{2}\mathbf{b}_1-\frac{5}{6}\mathbf{b}
_2+\frac{1}{12}\mathbf{b}_3\]\[
[\mathbf{b}_1,\mathbf{b}_3]=-5\mathbf{b}_1+\frac{11}{3}\mathbf{b}_2+\frac{5}{6}
\mathbf{b}_3-2J_\bbbi\mathbf{b}_1\]\[
[\mathbf{b}_2,\mathbf{b}_3]=-1653\mathbf{b}_1+5\mathbf{b}_2+\frac{5}{2}\mathbf{b
}_3+2J_\bbbi\mathbf{b}_2
\]
After an invertible transformation of the form (\ref{lintrans})
\[\hat{\mathbf{a}}_1=2\mathbf{b}_1,\qquad 
\hat{\mathbf{a}}_2=\frac{1}{6}\mathbf{b}_2,\qquad
\hat{\mathbf{a}}_3=\frac{5}{6}\mathbf{b}_1-\frac{5}{18}\mathbf{b}_2+\frac{1}{36}
\mathbf{b}_3,\qquad \hat{J}=\frac{1}{36}J_\bbbi+\frac{5}{12}\]
one can easily verify that the commutation relations for ${\bf 
\hat{a}_1,\hat{a}_2,\hat{a}_3}$ (structure constants) coincide with the ones 
for the $\bbbd_2$ group (\ref{com1}), and thus the two algebras are 
graded~isomorphic.
We treated all other cases similarly.\hfill $\blacksquare$

This covers the situation for all groups where we choose degenerate orbits, but 
we can also consider the case where we choose generic orbits (we did this for 
$\bbbd_2$ in Section 3.1). For all groups $\mathcal{G}$, we take the generators
\[\mathbf{a}_1 = \left<\frac{1}{\lambda-\mu}\sigma_1\right>_\mathcal{G},\quad
\mathbf{a}_2 = \left<\frac{1}{\lambda-\mu}\sigma_2\right>_\mathcal{G},\quad
\mathbf{a}_3 = \left<\frac{1}{\lambda-\mu}\sigma_3\right>_\mathcal{G}\]
with
\[\sigma_1=\begin{pmatrix}
0&1\\
1&0
\end{pmatrix},\quad\sigma_2=\begin{pmatrix}
0&-1\\
1&0
\end{pmatrix},\quad\sigma_3=\begin{pmatrix}
1&0\\
0&-1
\end{pmatrix}\]
being a basis of $sl(2,\mathbb{C})$. If $G$ is the M\"obius group
associated with the reduction group $\mathcal{G}$, the automorphic function $J$
is given by
\[J=\left<\frac{1}{\lambda-\mu}\right>_G\]
With all finite reduction groups the commutation relations take the form
\begin{eqnarray}\nonumber
&&[\mathbf{a}_1,\mathbf{a}_2]=p\mathbf{a}_1+q\mathbf{a}_2+r_3\mathbf{a}
_3+2J\mathbf{a}_3,\\ \label{genericc}
&&[\mathbf{a}_1,\mathbf{a}_3]=s\mathbf{a}_1+r_2\mathbf{a}_2-q\mathbf{a}
_3+2J\mathbf{a}_2,\\
&&[\mathbf{a}_2,\mathbf{a}_3]=r_1\mathbf{a}_1-s\mathbf{a}_2+p\mathbf{a}
_3+2J\mathbf{a}_1\nonumber
\end{eqnarray}
where $s$, $p$, $q$, $r_i$ are functions of a generic point $\mu$. For example, for the trivial group all coefficients $s$, $p$, $q$, $r_i$ are equal to zero,
for the group $\mathbb{Z}_N$ we have:
\[r_3=-\frac{2}{\mu},\quad p=q=s=r_1=r_2=0\]
and for $\mathbb{D}_N$ we have:
\[r_2=\frac{(\mu^{N}-1)^2}{\mu^{N+1}},\quad 
r_1=\frac{\mu^{2N}-1}{\mu^{N+1}},\quad s=p=q=r_3=0.\]
In fact using transformations (\ref{lintrans}) we can reduce the relations (\ref{genericc}) to one of the above listed cases.    

Here we assert that the list of automorphic Lie algebras  corresponding to all 
finite reduction groups and $\mathfrak{A}=sl(2,\mathbb{C})$ is rather short and 
up to graded isomorphism can be represented by algebras of the following 
types:
\begin{enumerate}
\item[$\mathcal{A}^0$] the polynomial part of the Loop algebra 
($\mathfrak{A}_{\lambda}(\infty)=\mathbb{C}[\lambda]\otimes_\mathbb{C} 
sl(2,\mathbb{C})$), when the reduction group is trivial;
\item[$\mathcal{A}^1$] the subalgebra $L_+(\mathfrak{A},\phi),\ \phi^2=id$ of 
the Kac-Moody algebra, which corresponds to 
$\mathfrak{A}_{\lambda}^\mathcal{G}(\Gamma)$ with 
$\mathcal{G}\simeq\mathbb{Z}_2$ and a degenerate orbit $\Gamma=\{\infty\}$;
\item[$\mathcal{A}^1_{1}$]  the algebra 
$\mathfrak{A}_{\lambda}^\mathcal{G}(\Gamma_1)$ with
$\mathcal{G}\simeq\mathbb{Z}_2$ and a generic orbit $\Gamma_1=\{\pm 1\}$;
\item[$\mathcal{A}^2$] the algebra $\mathfrak{A}_{\lambda}^\mathcal{G}(\Gamma)$ 
with $\mathcal{G}\simeq\mathbb{D}_2$ and a degenerate orbit 
$\Gamma=\{0,\infty\}$;
\item[$\mathcal{A}^2_{\mu}$] the algebra
$\mathfrak{A}_{\lambda}^\mathcal{G}(\Gamma_\mu)$  with
$\mathcal{G}\simeq\mathbb{D}_2$ and a
generic orbit $\Gamma_\mu$.
\end{enumerate}

\begin{Pro}\label{pro6} The algebras of the types
$\mathcal{A}^0$, $\mathcal{A}^1$, $\mathcal{A}^1_{1}$, $\mathcal{A}^2$ and 
$\mathcal{A}^2_{\mu}$ are not graded isomorphic.
\end{Pro}
{\bf Proof:} For any algebra $\mathcal{A}$ we set 
$\hat{\mathcal{A}}=\mathcal{A}/\langle[\mathcal{A},\mathcal{A}]\rangle$, where 
$\langle[\mathcal{A},\mathcal{A}]\rangle$ denotes the ideal generated by the 
commutator of the algebra $\mathcal{A}$ with itself. It follows from the 
commutation relations of these algebras, given earlier, that 
\[\mbox{dim}\hat{\mathcal{A}}^0=0,\quad  \mbox{dim}\hat{\mathcal{A}}^1=1,\quad  
\mbox{dim}\hat{\mathcal{A}}^1_{1}=2,\quad  
\mbox{dim}\hat{\mathcal{A}}^2=2,\quad  \mbox{dim}\hat{\mathcal{A}}^2_\mu=3.
\]

Our analysis of equations (\ref{P1P0}), (\ref{weq1}) and (\ref{weq2}), these 
being the equations determining the mapping between the bases of two algebras 
to establish a graded isomorphism between them, shows that the algebras 
$\mathcal{A}^1_\mu$ and $\mathcal{A}^2$ are not graded isomorphic, since in 
this case the equations (\ref{P1P0}), (\ref{weq1}) and (\ref{weq2}) have no 
solution.
\hfill  $\blacksquare$

\subsection{Explicit realisations of $sl(2,\bbbc)$ automorphic Lie algebras as finitely generated $\bbbc[J]$--Lie modules}\label{explicit}

Automorphic Lie algebras $\mathcal{A}^0$, $\mathcal{A}^1$, $\mathcal{A}^1_{1}$, $\mathcal{A}^2$ and 
$\mathcal{A}^2_{\mu}$  are almost graded infinite dimensional Lie  algebras over $\bbbc$. They also can be viewed as  $\bbbc[J]$--Lie modules with three generators $\ba_1, \ba_2, \ba_3$,  where $J$ is the corresponding automorphic function of the parameter $\lambda$ \cite{mik_dis}. Each of these algebras and the corresponding $\bbbc[J]$--Lie module can be extended by a derivation $\cD$.  In this Section we give explicit realisations for all automorphic Lie algebras listed above, their derivations and commutation relations.

$\mathcal{A}^0$: $\quad J=\lambda,\quad \cD=\dfrac{d}{d\,  \lambda},\quad
 \ba_1=\bbe,\quad \ba_2=\bbf,\quad \ba_3=\bbh,
$

with commutation relations (\ref{ls2}), $[\cD,\ba_i]=0,\ [\cD,J]=1$.

$\mathcal{A}^1$: $\quad J=\lambda^2,\quad \cD=\lambda \dfrac{d}{d\,  \lambda},\quad 
 \ba_1=\lambda \bbe,\quad \ba_2=\lambda \bbf,\quad \ba_3=\bbh,
$

and   commutation relations: $ [\cD,J]=2J$
\[ \begin{array}{lll}\phantom{}
   [\ba_1,\ba_2]=J\ba_3,\  & [\ba_3,\ba_1]= 2\ba_1,\  &  [\ba_3,\ba_2]= -2\ba_2,
   \\ \phantom{}
[\cD,\ba_1]=\ba_1,\ & [\cD,\ba_2]=\ba_2,\ & [\cD,\ba_3]=0.
 \end{array}
\]
$\mathcal{A}^1_{1}$: $\quad J=\dfrac{1}{\lambda^2-1},\quad \cD=\lambda \dfrac{d}{d\,  \lambda},
\quad  \ba_1=\dfrac{\lambda}{\lambda^2-1} \bbe,\quad \ba_2=\dfrac{\lambda}{\lambda^2-1} \bbf,\quad \ba_3=\bbh,$

with commutation relations: $[\cD,J]=-2J-2J^2$,
\[
 \begin{array}{lll}\phantom{}
[\ba_1,\ba_2]=(J+J^2)\ba_3,\  & [\ba_3,\ba_1]= 2\ba_1,\    &[\ba_3,\ba_2]= -2\ba_2,\\ \phantom{}
[\cD,\ba_1]=-(1+2J)\ba_1,\ & [\cD,\ba_2]=-(1+2J)\ba_2,\ & [\cD,\ba_3]=0.
 \end{array}
\]

$\mathcal{A}^2$: $\quad J=\lambda^2+\lambda^{-2},\quad \cD=\lambda (\lambda^2-\lambda^{-2}) \dfrac{d}{d\,  \lambda},$
\[
  \ba_1=\lambda \bbe+\lambda^{-1} \bbf,\quad \ba_2=\lambda^{-1} \bbe+\lambda \bbf,\quad \ba_3=2(\lambda^2-\lambda^{-2})\bbh,
\]
with commutation relations: $[\cD,J]=2J^2-8$,
\[
 \begin{array}{lll}\phantom{}
[\ba_1,\ba_2]=\ba_3,\  & [\ba_3,\ba_1]= 2J\ba_1-4\ba_2,\    &[\ba_3,\ba_2]= -2J\ba_2+4\ba_1, \\ \phantom{}
[\cD,\ba_1]=J\ba_1-2\ba_2,\ & [\cD,\ba_2]=J\ba_2-2\ba_1,\ & [\cD,\ba_3]=2J\ba_3.
 \end{array}
\]

$\mathcal{A}^2_{\mu}$ : $
J=\dfrac{(\mu^4-1)\lambda^2}{(\lambda^2-\mu^2)(1-\lambda^2\mu^2)},\quad \cD=\dfrac{\mu^2\lambda(1-\lambda^4)}{2(\lambda^2-\mu^2)(1-\lambda^2\mu^2)} \dfrac{d}{d\,  \lambda},
$
\[
 \ba_1=\dfrac{\lambda\mu}{\lambda^2-\mu^2} \bbe + \dfrac{\lambda\mu}{1-\lambda^2\mu^2}\bbf,\quad 
 \ba_2=\dfrac{\lambda\mu}{1-\lambda^2\mu^2}\bbe+\dfrac{\lambda\mu}{\lambda^2-\mu^2} \bbf  ,\quad \ba_3=\dfrac{\mu^2(\lambda^4-1)}{(\lambda^2-\mu^2)(1-\lambda^2\mu^2)}\bbh,
\]
with commutation relations: $[\cD,J]=J^3+2\alpha J^2+J$,
\[
 \begin{array}{ll}\phantom{}
[\ba_1,\ba_2]=J\ba_3,\  & [\cD,\ba_3]=(J^2+\alpha J)\ba_3,\\
\phantom{}
[\ba_3,\ba_1]= 2(J+\alpha)\ba_1+4\beta \ba_2,\    &
[\cD,\ba_1]=\dfrac{1}{2}(2J^2+3\alpha J+1)\ba_1+\beta J \ba_2,\\
\phantom{}
[\ba_3,\ba_2]= -2(J+\alpha)\ba_2-4\beta \ba_1, 
 &[\cD,\ba_2]=\dfrac{1}{2}(2J^2+3\alpha J+1)\ba_2+\beta J \ba_1,
 \end{array}
\]
where $\alpha=\dfrac{\mu^4+1}{\mu^4-1}$ and $\beta=\dfrac{\mu^2}{\mu^4-1}$.
\section{Integrable systems corresponding to
finite reduction groups}\label{sec4}

Automorphic Lie algebras can be used to find systems of integrable equations, 
by using them to construct a Lax pair $(L,A)$:
\begin{equation}\label{laxpair}
 L = \partial_x + {\bf U}(x,t,\lambda),\qquad
A = \partial_t + {\bf V}(x,t,\lambda)
\end{equation}
where ${\bf U}, {\bf V} \in \mathfrak{A}_{\lambda}^{\mathcal{G}}(\Gamma)$.

A Lax pair defines the linear differential system
\begin{equation}\label{linsys}
 L\psi=0,\quad A\psi=0
\end{equation}
where $\psi$ is a fundamental solution matrix to this linear problem. In order
for this linear system to be consistent, the following {\itshape compatibility
condition} must hold
\begin{equation}\label{compcond}
 {\bf V}_x - {\bf U}_t + [{\bf U},{\bf V}] = 0 ,
\end{equation}
which means that operators $L$ and $A$ commute $[L,A]=0$.

Let us denote as $\gA^{\cG}=\{a\in \gA\,|\, \Phi(a)=a, \forall \Phi\in \cG\}$   a
$\cG$-invariant subalgebra of a simple finite dimensional Lie algebra 
$\mathfrak{A}$. Let ${\bf G}^{\cG}$ be a Lie group corresponding to 
$\gA^{\cG}$ and ${\bf g}\in {\bf G}^{\cG}$ be a differentiable function of
$x,t$ with values in ${\bf G}^{\cG}$. 

\begin{Def}
 The map
\begin{equation}\label{gauge}
 L\to\hat{L}={\bf g}^{-1}L{\bf g},\qquad A\to\hat{A}={\bf g}^{-1}A{\bf g}
\end{equation}
is called  a gauge transformation of the Lax pair.
\end{Def}

Obviously 
\[\hat{L} = \partial_x + \hat{{\bf U}}(x,t,\lambda), \qquad \hat{A} = \partial_t
+ \hat{{\bf V}}(x,t,\lambda)\] 
where
\[\hat{{\bf U}}=\mathbf{g}^{-1}\mathbf{g}_x+\mathbf{g}^{-1}{\bf
U}\mathbf{g},\qquad \hat{{\bf
V}}=\mathbf{g}^{-1}\mathbf{g}_t+\mathbf{g}^{-1}{\bf V}\mathbf{g}\]
and $[\hat{L},\hat{A}]=0$. If $\psi$ is a fundamental solution of the problem
(\ref{linsys}), then $\chi={\bf g}^{-1}\psi$ is a fundamental solution of the
problem $\hat{L}\chi=0,\ \hat{A}\chi=0$.

 Lax
pairs $(L,A)$ and $(\hat{L},\hat{A})$  related by a gauge
transformation are called {\em gauge
equivalent}.  Choosing an appropriate gauge we can 
transform a Lax pair to a convenient form. 

\begin{Def}
 We say that a lax pair (\ref{laxpair}) is in the canonical gauge if ${\bf
U}\cap \gA^{\cG}=0$ and ${\bf V}\cap \gA^{\cG}=0$.
\end{Def}
The canonical gauge is almost unique. The remaining gauge freedom is due to
constant ($x,t$-independent) elements ${\bf g}\in {\bf G}^{\cG}$, which are
point symmetries of the resulting integrable non-linear system. 

There is also a freedom in the choice of independent variables $x,t$. Suppose
\begin{equation}\label{pointtrans}
 x=X(\xi,\eta),\qquad t=T(\xi,\eta)
\end{equation}
 is an invertible change of
variables, then 
\[ \psi_\xi=\frac{\partial X}{\partial \xi}\psi_x+\frac{\partial T}{\partial
\xi}\psi_t =-\frac{\partial X}{\partial \xi}{\bf U}\psi-\frac{\partial
T}{\partial\xi}{\bf V}\psi, \]
\[ \psi_\eta=\frac{\partial X}{\partial \eta}\psi_x+\frac{\partial T}{\partial
\eta}\psi_t =-\frac{\partial X}{\partial \eta}{\bf U}\psi-\frac{\partial
T}{\partial\eta}{\bf V}\psi \]
and thus in new variables the Lax pair corresponding to  (\ref{laxpair}) can be
written in the form
\[ \tilde{L}=\partial_\xi+\tilde{\bf U},\qquad 
\tilde{A}=\partial_\eta+\tilde{\bf V},\]
where 
\[
 \tilde{\bf U}=\frac{\partial X}{\partial \xi}{\bf U}+\frac{\partial
T}{\partial\xi}{\bf V},\quad  \tilde{\bf V}=\frac{\partial X}{\partial \eta}{\bf
U}+\frac{\partial
T}{\partial\eta}{\bf V}.
\]

In the following subsections we consider Lax pairs and corresponding second-order
systems of two equations using the automorphic Lie algebras constructed above.

\subsection{Equations corresponding to algebra $\mathcal{A}^2$}

Theorem 1 states that automorphic Lie algebras corresponding to the groups
$\mathbb{D}_N$, $\mathbb{T}$, $\mathbb{O}$ and $\mathbb{I}$ and degenerate
orbits are all graded isomorphic  and can be represented by algebra ${\cal
A}^2$.  For this algebra we choose a basis $\langle
\mathbf{a}_1,\mathbf{a}_2,\mathbf{a}_3, J \rangle$ with
commutation relations of the form (\ref{com1})
\[[\mathbf{a}_1,\mathbf{a}_2]=\mathbf{a}_3\quad [\mathbf{a}_1,\mathbf{a}_3]
=4\mathbf
{a}_2-2J\mathbf{a}_1\quad
[\mathbf{a}_2,\mathbf{a}_3]=-4\mathbf{a}_1+2J\mathbf{a} _2. \]
In order to obtain a system of two second order equations we choose the
following Lax pair
\begin{eqnarray}\label{Ldnls}
 L&=&\partial_x-\sum_{i=1}^3u_i(x,t)\mathbf{a}_i\\ \label{Adnls}
 A &=&\partial_{t}-\sum_{i=1}^3v_i(x,t)\mathbf{a}_i-\sum_{i=1}^3w_i(x,t)J\mathbf{a}_i
\end{eqnarray}
In this case the subalgebra $\gA^\cG$ is trivial  and therefore the
Lax pair is already in the canonical gauge.

Decomposing the compatibility condition (\ref{compcond}) over the basis we
obtain a system of eight equations
\begin{eqnarray}
 \mathbf{a}_1:&\quad & u_{1t}=v_{1x}-4(u_3v_2-u_2v_3)\label{8}\\
  \mathbf{a}_2:&\quad & u_{2t}=v_{2x}-4(u_1v_3-u_3v_1)\label{1}\\
 \mathbf{a}_3:&\quad & u_{3t}=v_{3x}-u_1v_2+u_2v_1\label{2}\\
 J\mathbf{a}_1:&\quad & 4(u_3w_2-u_2w_3)+2(u_3v_1-u_1v_3)-w_{1,x}=0\label{3}\\
 J\mathbf{a}_2:&\quad & 4(u_1w_3-u_3w_1)+2(u_2v_3-u_3v_2)-w_{2,x}=0\label{4}\\
 J\mathbf{a}_3:&\quad & u_1w_2-u_2w_1-w_{3,x}=0\label{5}\\
 J^2\mathbf{a}_1:&\quad & 2(u_3w_1-u_1w_3)=0\label{6}\\
 J^2\mathbf{a}_2:&\quad & 2(u_2w_3-u_3w_2)=0\label{7}.
\end{eqnarray}
for nine functions $u_i,v_i,w_i$. This system is underdetermined. In order to
make it well determined we use transformation (\ref{pointtrans}) of the form
$x\to X(x',t),\ t\to t$ to make $u_3=1$. Then it follows from
(\ref{5})-(\ref{7}) that $w_{3,x}=0$ and thus $w_3=w_3(t)$ is a function of $t$
only. By an appropriate change of the variable $t$, $t\to T(t')$, we can fix
$w_3=2$. We shall omit primes and use the notations $x$ and $t$ for the
new independent variables. Then from equations (\ref{2})-(\ref{7}) it follows
that 
\[
 w_1=2u_1,\ w_2=2u_2,\ v_3=-u_1u_2+\alpha,\ 
v_1= u_{1,x}-u_1^2u_2+\alpha u_1,\
v_2= -u_{2,x}-u_1u_2^2+\alpha u_2,\]
where $\alpha=\alpha(t)$ is an arbitrary function of $t$.

Finally, the equations at $\mathbf{a}_1$ and $\mathbf{a}_2$ give us our 
nonlinear integrable system:
\begin{equation}\label{dnls}
 \begin{array}{rcl}
u_{1,t}&=& u_{1,xx}-(u_1^2u_2)_x+4u_{2,x}+\alpha u_{1,x},\\
-u_{2,t}&=& u_{2,xx}+(u_1u_2^2)_x-4u_{1,x}-\alpha u_{2,x}.
 \end{array}
\end{equation}
The arbitrary function $\alpha(t)$ can be removed by a Galilean transformation $x\to x+\int \alpha(t)\,dt$.

There is a complete classification of second-order two component  integrable systems \cite{mr89e:58062}, 
\cite{mr89g:58092}. Equation (\ref{dnls}) corresponds to the equation $(D)$ in the list of integrable systems
provided in \cite{mr89g:58092}. This system is often called the deformed derivative nonlinear  Schr\"odinger equation. The well known derivative nonlinear Schr\"odinger equation \cite{DNLS} (equation (I) in \cite{mr89g:58092})
\[
  \begin{array}{rcl}
u_{1,t}&=& u_{1,xx}-(u_1^2u_2)_x ,\\
-u_{2,t}&=& u_{2,xx}+(u_1u_2^2)_x
 \end{array}
\]
corresponds to the algebra $\cA^1$ in a similar way. The Lax representations for the famous nonlinear Schr\"odinger equation and the Heisenberg model originate from the algebra $\cA^0$. They  correspond to different choices of the gauge of the Lax operator.

System (\ref{dnls}) possesses an infinite hierarchy of symmetries. They can be found using the same $L$ operator (\ref{Ldnls}) and $A_k,\ k\in\bbbn$ operators of the form 
\[
 A_k=\partial_{t_k}-\sum_{s=1}^k\sum_{i=1}^3 v^{(s)}_iJ^{s-1}\mathbf{a}_i.
\]
In particular (\ref{dnls}) corresponds to $k=2$
\[
 t=t_2,\qquad  A=A_2=\partial_{t_2}-\sum_{i=1}^3(v_i\mathbf{a}_i +
 2u_iJ\mathbf{a}_i)
\]
 and in the case of $k=1$ the system is linear
\[
 (u_1)_{t_1}=u_{1,x},\qquad (u_2)_{t_1}=u_{2,x}.
\]
The coefficients $v^{(s)}_i$ can be found from the compatibility condition $[L,A_k]=0$, or using the method proposed in \cite{mr86h:58071}, \cite{sokbook}.

The existence of the derivation 
$$\cD=\dfrac{\mu^2\lambda(1-\lambda^4)}{2(\lambda^2-\mu^2)(1-\lambda^2\mu^2)} \dfrac{d}{d\,  \lambda}$$ 
of the automorphic Lie algebra $\cA^2$ enables us  to construct a Lax representation for the master symmetry. Let us consider the same Lax operator $L$ (\ref{Ldnls}) and define the second operator $M$, which includes the $\cD$ derivation 
\[
 M=\dfrac{\partial}{\partial\tau}+\cD-\sum_{i=1}^3(V_i\mathbf{a}_i +
W_iJ\mathbf{a}_i).
\]
It follows from the compatibility conditions $[L,M]=0$ that $W_i=2x u_i+\gamma u_i$, where $\gamma $ is an arbitrary function of $\tau$. We choose $\gamma=0$ without loss of generality. Then we can establish that 
\[
 V_1=\frac{1}{2} u_1+ u_1 V_3+ x u_{1,x},\qquad V_2=-\frac{1}{2} u_2+  u_2 V_3- x u_{2,x},\qquad 
V_3=-x u_1 u_2+\alpha.
\]
We shall omit the inessential constant of integration $\alpha$ (an arbitrary function of $\tau$). The resulting system
\begin{equation}\label{msym}
\begin{array}{l}
 u_{1,\tau}=4 u_2-u_1^2 u_2+\dfrac{3}{2} u_{1,x}+x (u_{1,x}-u_{1}^2 u_{2}
+4 u_{2})_x,\\ \\
 u_{2,\tau}=4 u_{1}-u_{1} u_{2}^2-\dfrac{3}{2} u_{2,x}-x (u_{2,x}+u_{2}^2 u_{1}
-4 u_{1})_x
  \end{array}
\end{equation}
is a master symmetry of the the system (\ref{dnls}). Indeed, $\partial_t$ and $\partial _\tau$  do not commute, but their commutator commutes with $\partial_t$ and defines a symmetry of (\ref{dnls}):
\begin{equation}\label{dnls3}
\begin{array}{l}\phantom{}
 \partial_{t_3}u_1=[\partial_\tau,\partial_t]u_1=2 u_{1,xxx}+(16 u_1-4 u_1^3-12 u_1 u_2^2+3 u_1^3 u_2^2-6 u_1 u_2 u_{1,x})_x,\\ \phantom{}
 \partial_{t_3}u_2=[\partial_\tau,\partial_t]u_2=2 u_{2,xxx}+(
16 u_2-4 u_2^3-12 u_1^2 u_2+3 u_1^2 u_2^3+6 u_1 u_2 u_{2,x})_x.
  \end{array}
\end{equation}
An infinite hierarchy of commuting local symmetries of equation (\ref{dnls}) can be constructed recursively 
\[\partial_{t_{n+1}}=[\partial_\tau,\partial_{t_n}],\qquad [\partial_{t_n}\partial_{t_m}]=0.\] 
Moreover, the operators $A_k$ can also be found recursively
$
 A_{k+1}=[M,A_k].
$

For example, taking
\[
 \begin{array}{ll}
  A_2=\partial_{t_2}-\cV,\qquad &\cV=\sum_{i=1}^3(v_i+2u_iJ)\mathbf{a}_i,\\
  M=\partial_{\tau}+\cD-\cW,&\cW=\sum_{i=1}^3(W_i+2xu_iJ)\mathbf{a}_i,
 \end{array}
\]
we obtain
\[
 [M,A_2]=[\partial_{\tau},\partial_{t_2}]-\cV_\tau-\cD(\cV)+\cW_{t_2}+[\cW,\cV]=\partial_{t_3}-\sum_{i=1}^3(z_i+4v_iJ+8u_iJ^2)\mathbf{a}_i
\]
where
\[
\begin{array}{l}
z_1= 2 u_{1, xx }-6 u_2 u_1 u_{1,x}+8 u_{2,x}+3 u_2^2 u_1^3-4 u_1^3-4 u_2^2 u_1-16 u_1,
\\
z_2= 2 u_{2, xx }+6 u_1 u_2 u_{2,x}-8 u_{1,x}+3 u_1^2 u_2^3-4 u_2^3-4 u_1^2 u_2-16 u_2,
\\
z_3= 2 u_1 u_{2,x}-2 u_2 u_{1,x}+3 u_2^2 u_1^2-4 u_1^2-4 u_2^2-16
.
\end{array}
\]
The Lax pair $(L,A_3)$ yields equations (\ref{dnls3}).

\subsection{Equations associated with generic orbits of  $sl(2,\bbbc)$ automorphic Lie algebras}

In the case of $sl(2,\bbbc)$ the automorphic Lie algebras $\gA^\cG_\lambda(\Gamma_\mu)$ corresponding to a generic orbit of a finite group have a basis $\langle 
\mathbf{a}_1,\mathbf{a}_2,\mathbf{a}_3, J \rangle$ such that the commutation 
relations take the form
\[[\mathbf{a}_1,\mathbf{a}_2]=p\mathbf{a}_1+q\mathbf{a}_2+r_3\mathbf{a}
_3+2J\mathbf{a}_3\]
\[[\mathbf{a}_1,\mathbf{a}_3]=s\mathbf{a}_1+r_2\mathbf{a}_2-q\mathbf{a}
_3+2J\mathbf{a}_2\]
\[[\mathbf{a}_2,\mathbf{a}_3]=r_1\mathbf{a}_1-s\mathbf{a}_2+p\mathbf{a}
_3+2J\mathbf{a}_1\]
where $s$, $p$, $q$, $r_i$  are constants depending on a generic 
point $\mu$ and the choice of representation of a reduction group $\cG$. The algebra  $\gA^\cG_\lambda(\Gamma_\mu)$ is almost graded
\[
  \gA^\cG_\lambda(\Gamma_\mu)=\bigoplus_{k=1}^\infty \gA^k,\qquad [\gA^p,\gA^q]\subset \gA^{p+q}\oplus\gA^{p+q+1}
\]
where $\gA^{1}={\rm span} _\bbbc\langle \mathbf{a}_1,\mathbf{a}_2,\mathbf{a}_3\rangle$ and $\gA^{k}=J^{k-1} \gA^{1}$.

Let us take a 
 Lax pair $(L,A)$ where the operator $L$ is spanned by the first homogeneous space $\gA^1$ of the automorphic Lie algebra, while the operator $A$ is spanned by the first and second homogeneous spaces:
\[L=\partial_x+\sum_{i=1}^3S_i(x,t)\mathbf{a}_i\]
\[A=\partial_t+\sum_{i=1}^3v_i(x,t)\mathbf{a}_i+\sum_{i=1}^3w_i(x,t)J\mathbf{a}
_i \ .\] 
The compatibility condition $[L,A]=0$ results in a system of equations in the first three homogeneous spaces. In $\gA^3$ we get the equations (vanishing the coefficients at $J^2\mathbf{a}_1,J^2\mathbf{a}_2,J^2\mathbf{a}_3$ respectively):
\[
 2(S_2w_3 - S_3w_2) = 0,\quad 
2(S_1w_3 -S_3w_1) = 0,\quad 
2(S_1w_2 - S_2w_1) = 0.
\]
and thus $w_i=\gamma(x,t)S_i$. Taking this into account we see that the coefficients at $J\mathbf{a}_1,J\mathbf{a}_2,J\mathbf{a}_3$ vanish if
\begin{equation}\label{veq}
  w_{1,x} + 2(S_2v_3 -S_3v_2) = 0,\quad
w_{2,x} + 2(S_1v_3 -S_3v_1)= 0,\quad
w_{3,x} + 2(S_1v_2 -S_2v_1)= 0.
\end{equation}
Equations (\ref{veq}) are compatible and enable us to 
 express the functions $v_i$ in terms of $S_i$ and their $x$--derivatives if
\[
S_1w_{1,x} + S_2w_{2,x} + S_3w_{3,x} = \gamma_x(S^2_1- S^2_2+ S^2_3) +\frac{1}{2}\gamma(S^2_1- S^2_2+ S^2_3)_x = 0.
\]
Assuming that $S^2_1- S^2_2+ S^2_3\not\equiv 0$ we can make a change of variables $x\to \hat{x}=\alpha(x,t),\ t\to \hat{t}=\beta(t)$ (and thus $S_i\to \hat{S}_i=S_i/\alpha_x$), such that $\hat{S}^2_1- \hat{S}^2_2+ \hat{S}^2_3=1$ and $\gamma=2$. 
Thus, without a loss of generality we shall assume that 
\[
 S_1^2-S_2^2+S_3^2=1,\qquad w_i=2S_i.
\]
Taking this into account we represent a general solution of eq (\ref{veq}) in the form
\[
 v_1 =S_{2,x} S_3-S_{3,x} S_2  + \Phi S_1 ,\quad 
 v_2 =S_{1,x} S_3-S_{3,x} S_1  + \Phi S_2 ,\quad 
 v_3 =S_{1,x} S_2-S_{2,x} S_1  + \Phi S_3 ,\quad 
\]
where $\Phi=\Phi(x,t)$ is an as yet undetermined function. 
In $\gA^1$ the coefficients at $\mathbf{a}_1,\mathbf{a}_2,\mathbf{a}_3$ vanish if 
\[\begin{array}{l}
S_{1,t} = v_{1,x} + p(S_1v_2 - S_2v_1) + s(S_1v_3 - S_3v_1) + r_1(S_2v_3 - S_3v_2),\\
S_{2,t} = v_{2,x} + q(S_1v_2 - S_2v_1) + r_2(S_1v_3 - S_3v_1) - s(S_2v_3 - S_3v_2),\\
S_{3,t} = v_{3,x} + r_3(S_1v_2 - S_2v_1) - q(S_1v_3 - S_3v_1) + p(S_2v_3 - S_3v_2).
  \end{array}
\]
It follows from the equation $(S_1^2-S_2^2+S_3^2)_t=0$ that
\[
 \Phi = pS_1S_3 + sS_1S_2 - qS_2S_3+\frac{1}{2}
(r_1S^2_1- r_2S^2_2+ r_3S^2_3)+\theta(t)
\]
where $\theta(t)$ is an arbitrary function. Finally we obtain the following integrable system of equations 
\begin{eqnarray*}
 S_{1,t}& =&S_3S_{2,xx} - S_2S_{3,xx} + [S_1(pS_1S_3 + sS_1S_2 - qS_2S_3)]_x\\
&+&
\frac{1}{2}[S_1(r_1S^2_1- r_2S^2_2+ r_3S^2_3)]_x - pS_{3,x} - sS_{2,x} - r_1S_{1,x} + \theta(t) S_{1,x},\\
S_{2,t}& =&S_3S_{1,xx} - S_1S_{3,xx} + [S_2(pS_1S_3 + sS_1S_2 - qS_2S_3)]_x\\
&+&
\frac{1}{2}[S_2(r_1S^2_1- r_2S^2_2+ r_3S^2_3)]_x - qS_{3,x} -r_2S_{2,x} +sS_{1,x} + \theta(t) S_{2,x},\\
S_{3,t}& =&S_2S_{1,xx} - S_1S_{2,xx} + [S_3(pS_1S_3 + sS_1S_2 - qS_2S_3)]_x\\
&+&
\frac{1}{2}[S_3(r_1S^2_1- r_2S^2_2+ r_3S^2_3)]_x - r_3S_{3,x} +qS_{2,x} - pS_{1,x} + \theta(t) S_{3,x}.
\end{eqnarray*}
The functions $S_1(x,t),S_2(x,t),S_3(x,t)$ satisfy the condition $ S_1^2-S_2^2+S_3^2=1$ and can be parametrised by two functions $u=u(x,t)$ and $v=v(x,t)$:
\[S_1=\frac{1-uv}{u-v},\quad S_2=\frac{1+uv}{u-v},\quad S_3=\frac{u+v}{u-v}.\]
In these new variables the above system takes the form
\[\begin{array}{rcl}
u_t&=&u_{xx}-\dfrac{2u_x^2}{u-v}-\dfrac{2}{(u-v)^2}[2P(u,v)u_x-P(u,u)v_x]+\eta (t)u_x\\
-v_t&=&v_{xx}-\dfrac{2v_x^2}{u-v}+\dfrac{2}{(u-v)^2}[2P(u,v)v_x-P(v,v)u_x]-\eta (t)v_x
  \end{array}\]
where
\[
P(u,v)=2au^2v^2+b(uv^2+vu^2)+2cuv+d(u+v)+2e,\quad \eta (t)=\theta(t)-(r_1+r_2-r_3)/2
\]
and
\[a=\frac{1}{8}(r_2-r_1+2s),\quad b=\frac{1}{2}(p+q),\quad 
c=\frac{1}{4}(r_2+r_1-2r_3),\quad d=\frac{1}{2}(q-p),\quad e=\frac{1}{8}(r_2-r_1-2s).\]
The function $\eta (t)$  can be set to zero by the Galilean transformation $x\to x+\int \eta (t)\, dt$.
The system obtained corresponds to the system $(m)$ in the list in \cite{mr89g:58092}. In the simplest case of vanishing constants $s=p=q=r_i=0$ the system is equivalent (up to invertible point transformations) to the Heisenberg model, gauge equivalent to the nonlinear Schr\"odinger equation and the corresponding algebra is $\cA^0$. 

\section*{Acknowledgments} The authors are grateful to the Ministry of Science and Higher Education of the Russian Federation (agreement  075-02-2020-1514) and the EPSRC grant EP/P012655/1 for  partial support.


\end{document}